\documentclass[reprint,amsmath,amssymb,aps,superscriptaddress]{revtex4-1}

\usepackage{graphicx}
\usepackage{dcolumn}
\usepackage{bm}
\usepackage{braket}
\usepackage{color}
\usepackage{pstricks,pst-grad,color}
\usepackage{appendix}
\usepackage{here}
\usepackage{url}
\usepackage{hyperref}
\usepackage{comment}
\usepackage{mathtools}

\newcommand{\m}[1]{\mathcal{#1}}
\newcommand{\mr}[1]{\mathrm{#1}}

\newcommand{\bk}{\bm{k}}
\newcommand{\p}{\partial}
\newcommand{\II}{I\hspace{-.01em}I}

\usepackage{tikz}
\usetikzlibrary{shapes,calc}
\usetikzlibrary{shapes.geometric,arrows.meta,decorations.markings}

\begin{document}
\preprint{APS}

\title{Dissipation and geometry in nonlinear quantum transports of multiband electronic systems}
\author{Yoshihiro Michishita}
\email{yoshihiro.michishita@riken.jp}
 \affiliation{RIKEN Center for Emergent Matter Science (CEMS), Wako, Saitama 351-0198, Japan}
\author{Naoto Nagaosa}%
 \email{nagaosa@scphys.kyoto-u.ac.jp}
 \affiliation{RIKEN Center for Emergent Matter Science (CEMS), Wako, Saitama 351-0198, Japan}
 \affiliation{Department of Applied Physics, The University of Tokyo, Bunkyo, Tokyo 113-8656, Japan}

\date{\today}

\begin{abstract}
Nonlinear responses in condensed matters attract recent intensive interest because they provide rich information about the materials and hold the possibility of being applied in diodes or high-frequency optical devices.
Nonlinear responses are often closely related to the multiband nature of the system which can be taken into account by the geometric quantities such as the Berry curvature as shown in  the nonlinear Hall effect. Theoretically, the semi-classical Boltzmann treatment or the reduced density matrix method have been often employed, in which the effect of dissipation is included through the relaxation time approximation. In the diagrammatic method, the relaxation is treated through the imaginary part of the self-energy of the Green function and the consequent broadening of the spectral function for the integration over the real frequency. 
Therefore, the poles of the Green function do not play explicit pole when there is finite dissipation. 
In this paper, in stark contrast to this conventional picture, we show that the poles of the Green function determine mostly the nonlinear response functions with dissipation, which leads to the terms with the Fermi distribution function of complex argument and results in the dissipation-induced geometric term. We elucidate the geometric origin of the nonreciprocal conductivity, which is related to the Berry curvature generalized to the higher derivative. Finally, we derive the analytical results on the geometric terms of the nonlinear conductivities in the type-I and type-\II \ Weyl Hamiltonian to demonstrate their crucial roles.
\end{abstract}

\maketitle

\section{Introduction}

Recently, nonlinear response in the bulk systems in condensed matter physics has been intensively studied, especially in the context of the higher-harmonic generations\cite{Petersen2006,Zhao2016,Harter295,PhysRevB.48.11705,Morimotoe1501524,Wu2017}, nonlinear Hall effect\cite{PhysRevLett.115.216806,Ma2019,Kang2019,Dzsabere2013386118,PhysRevApplied.13.024053}, photovoltaic effect\cite{McIver2012,Kastl2015,PhysRevB.95.041104,Cook2017,Isobeeaay2497}, and non-reciprocal transport\cite{doi:10.1063/1.107968,Morimoto2018,Tokura2018,PhysRevB.99.115201,Wakatsukie1602390,Itahashieaay9120,Ando2020,yuan2021supercurrent,daido2021intrinsic,he2021phenomenological} because they have the information on the symmetry of material and have the possibility of the application to devices. For example, we can detect the parity-breaking of the bulk through the detection of the second harmonic generation\cite{Petersen2006,Zhao2016,Harter295}. The photovoltaic effect in bulk has the possibility to the high-frequency rectification devices\cite{Isobeeaay2497}. Moreover, the large non-reciprocality was found in the superconductors\cite{Wakatsukie1602390,Itahashieaay9120,Ando2020} and can be applied to the diode devices. 

In the linear conductivity, the effect of dissipation and multi-band contribution are usually separated, such as the anomalous quantum Hall effect, which can be described by the Berry curvature and is not affected by dissipation. 
On the other hand, in the nonlinear conductivity, both of dissipation and multi-band contribution intertwine the novel transport. For examples, the Berry curvature dipole term is proportional to the lifetime, which is the inverse of the strength of dissipation, while it is also proportional to the Berry curvature\cite{Du2019,PhysRevB.100.195117,Du2021}. Moreover, it has been pointed out in Ref.\cite{Morimoto2018} that, for the nonreciprocal current under the time-reversal symmetry (TRS), both dissipation and multi-band are necessary. Therefore, it is essential to analyze properly the effect of dissipation on the multi-band contribution in the nonlinear conductivities.

In most previous studies, the nonlinear response has been commonly studied by the semi-classical Boltzmann (SCB) treatment\cite{PhysRevLett.115.216806,PhysRevB.100.195117,PhysRevB.94.245121,Du2019} or the reduced density matrix (RDM) method\cite{PhysRevB.48.11705,PhysRevB.96.035431,PhysRevB.97.235446,PhysRevX.11.011001,watanabe2021nonreciprocal1,watanabe2021nonreciprocal2}. In these methods, it is not easy to consider the effect of dissipation rigorously, and therefore, we usually use the relaxation time approximation (RTA) to include the dissipation, or calculate each relaxation time for various scattering such as the side-jump and skew scattering\cite{PhysRevB.100.195117,Du2019}. RTA has the problem that it breaks the gauge invariance between the velocity gauge and the length gauge\cite{PhysRevB.96.035431,PhysRevB.103.195133}, and it cannot describe the proper relaxation when considering the finite input frequency\cite{PhysRevB.103.195133},
while it well describes the relaxation when considering the DC input\cite{PhysRevB.103.195133}.

In the microscopic diagrammatic theory, the relaxation is treated by the imaginary part of the self-energy of the Green’s function and the vertex corrections. The former also leads to the broadening of the spectral function, which appears in the integration over the real frequency. The reason for this integration path is to avoid the poles of the Fermi distribution function at Matsubara frequencies, but those contributions are small for nonlinear responses as will be shown below. Therefore, the poles of the Green’s with imaginary parts and the Fermi distribution function with complex argument plays an important role. 
In the previous studies, while the relaxation of the non-equilibrium states can be considered through the RTA, the effect of the broadening of the distribution function cannot be captured.
 
In this paper, we analyze the effect of the broadening of the spectral function on the nonlinear transport. We elucidate that the broadening of the spectral function results in the shift of the Fermi distribution function (DF) to the imaginary direction and the Matsubara term, which cannot be described by the SCB treatment or the RDM method. The shift of the Fermi FD to the imaginary direction gives the novel dissipation-induced geometric terms. For example, the Christoffel symbol term appears in second-order nonlinear transport, and this term also gives the multi-band correction to the nonlinear Drude term. We note that this Christoffel symbol term is completely different from Ref.\cite{PhysRevLett.112.166601} and Ref.\cite{PhysRevLett.122.227402}, in which it appears under the magnetic field or the nonuniform electric field. We also elucidate the geometric origin of the non-reciprocal conductivity, which is related to the Berry curvature generalized to higher derivative and is also a dissipation-induced geometric term.
Moreover, we analytically derive the geometric term in the Weyl Hamiltonian both for the type-I and type-I\hspace{-.01em}I cases. We show that the chemical potential dependence of the nonlinear Hall conductivity for each case is completely different, and therefore, the observation can be the detection of the Weyl points and its type. Especially for the type-I case, we also show that we can estimate the relaxation time in the material from this observation.
 
In the following, we derive the shift of the Fermi DF to the imaginary direction and the broadening terms from the Green function methods\cite{PhysRevB.99.045121,PhysRevB.103.195133}. First, we derive them in the linear conductivity to illustrate the formulation in section~\ref{Sec:linear}. We show the shift of the Fermi DF results in the quantum metric term at the Fermi surface. 
In section~\ref{Sec:nonlinear}, we extend the results in the linear response to the nonlinear transport, and derive the geometrical terms, such as the Christoffel symbol term and the generalized Berry curvature term. Then, we numerically calculate it in a model for transition metal dichalcogenides and show its dissipation-strength and chemical potential dependence.
In section~\ref{Sec:Weyl}, we derive the analytical results of the geometric terms such as the Berry curvature dipole term and the Christoffel symbol term for the type-I and type-I\hspace{-.01em}I Weyl Hamiltonian.
In section~\ref{Sec:Summary}, we summarize our results.
 
\section{Dissipative geometry in linear conductivity\label{Sec:linear}}
Before considering nonlinear conductivity, we first analyze dissipation effect in linear conductivity. Although the methods we use in this paper is not so effective in linear conductivity, the analysis in linear conductivity is pedagogical and helps us to understand the results in nonlinear conductivity.
\subsection{Formulation}
In this paper, we include the dissipation effect via the imaginary part of the single-particle self-energy and calculate the conductivity with the Green function methods. Throughout this paper, we ignore the momentum and frequency dependence of the dissipation and suppose the dissipation strength is same for all bands. This assumption and approximation are justified when we consider the impurity scattering, which is independent of the momentum transfer under the Born approximation. It is also justified to ignore the vertex correction because here we ignore the momentum dependence of the self-energy and satisfy the Ward-Takahashi equation. Under these approximations and assumptions, the single-particle Green function has the same eigenstate as the Hermitian part of the effective Hamiltonian $\m{H}_{\mr{eff}}=\m{H}_0 + \mr{Re}\Sigma^R$. We also set $e=k_B=\bar{h}=1$ throughout this paper.

\begin{figure}[t]
    \centering
    \includegraphics[width=0.98\linewidth]{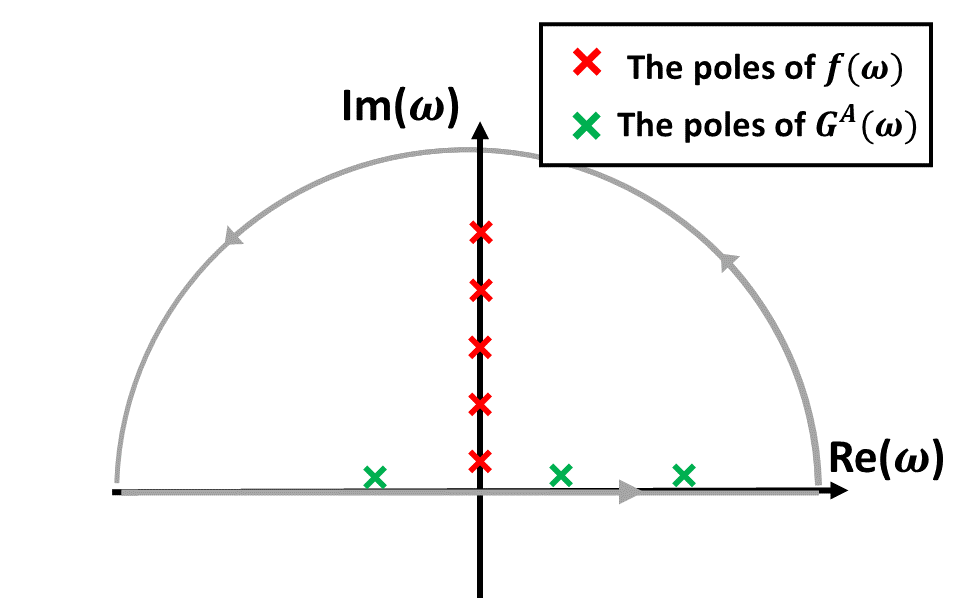}
    \caption{Path integration \\
    Path integration in the DC limit.} 
    \label{fig:Path}
\end{figure}

First, we analyze the dissipation effect through the distribution function in the linear conductivity, and focus on the symmetric part of the linear DC conductivity $\sigma^{\alpha\beta}_{DC} = (\sigma^{\alpha;\beta} + \sigma^{\beta;\alpha})/2$ for simplicity. $\alpha(\beta)$ in $\sigma^{\alpha;\beta}$ represents the output(input) direction. 
$\sigma^{\alpha\beta}_{DC}$ can be written in the Green function methods with the band-indices $n,m$ as,
\begin{eqnarray}
   &&\sigma^{\alpha\beta}_{\mr{DC}}\nonumber\\
   &&= \sum_{\bk}\int_{-\infty}^{\infty}\frac{d\omega}{2\pi}\mr{Re}\sum_{nm}\m{J}^{\alpha}_{nm}G^R_m\m{J}^{\beta}_{mn}\bigl(G^R_n-G^A_n\bigr)\frac{\p f}{\p\omega}\label{Green_linear}
\end{eqnarray}
where $\m{J}^\alpha = \p^{\alpha}\m{H}_{\mr{eff}}$, $\p^{\alpha} = \p/\p k^{\alpha}$,  $\m{O}_{nm}=\braket{n|\m{O}|m}$, $\m{O}_n=\m{O}_{nn}$, $\ket{n}$ is the eigenstates of $\m{H}_{\mr{eff}}$, and $f(\omega)$ is the Fermi distribution function. Throughout this paper, we omit writing the momentum dependence of the function, such as $\m{J}, \epsilon_n, G^{R(A)}_n$, and the frequency dependence of the Green function. In the limit $|\omega|\rightarrow\infty$, the integrand is proportional to $1/|\omega|^3$, and therefore, the integration $\int_{-\infty}^{\infty}d\omega/2\pi$ is equivalent to the contour integral $\oint_{C}d\omega/2\pi$.(along the closed loop $C$ in Fig.~\ref{fig:Path}.) For this integral, we should consider the poles of the advanced Green function (green cross marks in Fig.~\ref{fig:Path}.) and the Matsubara frequencies from the Fermi distribution function (red cross marks in Fig.~\ref{fig:Path}.). Then, Eq.~(\ref{Green_linear}) can be written as,
\begin{eqnarray}
   \sigma^{\alpha\beta}_{\mr{DC}} &=& \sigma^{\alpha\beta}_{\mr{M}} +\sigma^{\alpha\beta}_{\mr{G}} \\
   \sigma^{\alpha\beta}_{\mr{M}}&=&\mr{Re}\sum_{\bk}\frac{i}{2\beta}\sum_{nm}\sum_{\omega_M>0}(\m{J}^{\alpha}_{nm}\m{J}^{\beta}_{mn}+\m{J}^{\beta}_{nm}\m{J}^{\alpha}_{mn})\nonumber\\
   && \ \ \ \ \ \ \ \ \ \ \ \times\Bigl[\frac{\p}{\p\omega}\Bigl(G^R_m(G^R_n-G^A_n)\Bigr)\Bigr]_{i\omega_M}\label{polesM}\\
   \sigma^{\alpha;\beta}_{\mr{G}}&=& \mr{Re}\sum_{\bk}\sum_n\Bigl[\m{J}^{\alpha}_n\m{J}^{\beta}_n\tau\nonumber\\
   && \ +i\sum_{m\neq n}\frac{(\m{Q}^{\alpha\beta}_{D;n,m}\!+\!\m{Q}^{\beta\alpha}_{D;n,m})}{2}(\epsilon_{nm}\!+\!2i\eta) \Bigr]\Bigl(-\frac{\p f}{\p\omega}\Bigr)_{\epsilon_n+i\eta},\nonumber\\
   &&\label{polesA}
\end{eqnarray}
with
\begin{eqnarray}
   && \m{Q}^{\alpha\beta}_{D;n,m} =  \frac{\m{J}^{\alpha}_{nm}\m{J}^{\beta}_{mn}}{(\epsilon_{nm} + 2i\eta)^2},
\end{eqnarray}
where $\omega_M = (2M+1)\pi/\beta$ is the Fermionic Matsubara frequency, $\beta$ is the inverse of temperature, $\epsilon_n$ is the eigenvalue of $\m{H}_{\mr{eff}}$,$\eta = -\mr{Im}\Sigma^R$, $\tau = 1/(2\eta)$, $\epsilon_{nm}=\epsilon_n-\epsilon_m$ and $\m{Q}^{\alpha\beta}_{D;n} \equiv \sum_{m\neq n}\m{Q}^{\alpha\beta}_{D;n,m}$ is the dissipative quantum geometric tensor, which correspond to $\braket{\p^{\alpha}n|\p^{\beta} n}$ in the limit of $\eta\rightarrow0$.
We call the first term $\sigma^{\alpha;\beta}_M$ as `` Matsubara term'' because it represents the contribution from the poles of the Fermi distribution functions at the Fermionic Matsubara frequencies. 
While the contribution from the poles of the advanced Green function $\sigma_G$ corresponds to the results by the SCB treatment or the RDM method in $\eta\rightarrow0$ limit, the Matsubara term $\sigma_M$ is proportional to $\eta$\footnote{We can see $\sigma_M\propto\eta$ from $(G^R_n-G^A_n)_{i\omega_M}=2i\eta/((i\omega_M-\epsilon_n)^2+4\eta^2)$. We can also see $\sigma_M\propto\eta$ in figure~\ref{fig:mudep}.} and therefore cannot be condsidered in the SCB treatment or the RDM method.
Next, we analyze the second term in Eq.~(\ref{polesA}). 
We can understand the dissipative geometric term as the multi-band correction to the Drude term, which reads,
\begin{eqnarray}
       \tilde{\sigma}^{\alpha\beta}_{\mr{Drude}}&=& \sigma^{\alpha\beta}_{\mr{Drude}} +\sigma^{\alpha\beta}_{\mr{QM:re}}\\ \sigma^{\alpha\beta}_{\mr{QM:re}}&\sim&\sum_{\bk}\sum_n\frac{g^{\alpha\beta}_{S;n}}{\tau} \Bigl(-\frac{\p f}{\p\omega}\Bigr)_{\epsilon_n}\label{L_BI_QMre}
\end{eqnarray}
\begin{eqnarray}
       g^{\alpha\beta}_{S;n}&=&\sum_m g^{\alpha\beta}_{S;n,m}, \ \ g^{\alpha\beta}_{S;n,m} = \frac{(\m{J}^{\alpha}_{nm}\m{J}^{\beta}_{mn}+\m{J}^{\beta}_{nm}\m{J}^{\alpha}_{mn})}{2(\epsilon^2_{nm} + 4\eta^2)}.\nonumber\\
\end{eqnarray}
The detail derivation is written in appendix~\ref{app:Linear}. We can derive $\sigma^{\alpha\beta}_{\mr{QM:re}}$ from the dissipative geometric term.
We call $g^{\alpha\beta}_{S;n}$ as the ``smeared quantum metric,'' which becomes the quantum metric in $\eta\rightarrow0$ limit, and $\sigma^{\alpha\beta}_{\mr{QM:re}}$ as the ``(smeared) quantum metric term''. At the band degeneracy where $\epsilon_{nm}=0$, the smeared quantum metric is proportional to $\tau^2$, and therefore, the quantum metric term becomes proportional to $\tau$. This means the quantum metric term describes the multi-band correction to the Drude term which is proportional to $\tau$.
We note that we can also derive this correction from the reduced density matrix methods under the RTA, by changing $f(\epsilon_n)\rightarrow f(\epsilon_n + i\eta)$ and approximating $\mr{Im}f(\epsilon_n+i\eta)\simeq \eta(\p f/\p\omega)_{\epsilon_n}$. Therefore, we can consider the quantum metric term stems from the imaginary part of the Fermi DF. We also numerically check how large these terms and show that, the quantum metric term are dominant when the band degeneracy exists at the Fermi surface in appendix~\ref{app:linear_num}. We also show in appendix~\ref{app:linear_num} that the Matsubara term is also finite, and therefore, the treatment in this section is not so effective in linear response cases.
On the other hand, as we will show in the next section, the Matsubara term is small and the description by the shift of the Fermi DF to the imaginary direction works well in the nonlinear responses. 

\section{Dissipation-induced geometry in nonlinear response\label{Sec:nonlinear}}
\subsection{formulation}
Next, we consider the effect of the broadening of DF on second-order nonlinear DC conductivity by the same procedure as the linear case. In nonlinear conductivity, a lot of terms emerge from the dissipation effect and the formula becomes so complicated. Therefore, we write the detail derivation in appendix~\ref{app:NLR},  and here write down the final and summarized results, which read,
    \begin{eqnarray}
       \sigma^{\alpha;\beta\gamma}_{\mr{DC}} &=& \sigma^{\alpha;\beta\gamma}_{\mr{M}} + \sigma^{\alpha;\beta\gamma}_{\mr{Drude}} + \sigma^{\alpha;\beta\gamma}_{\mr{BCD}} + \sigma^{\alpha;\beta\gamma}_{\mr{ChS}} + \sigma^{\alpha;\beta\gamma}_{\mr{gBC}} + \m{O}(\tau^{-2}).\nonumber\\
        \label{NLR}
    \end{eqnarray}
$\sigma^{\alpha;\beta\gamma}_{\mr{M}}$, $\sigma^{\alpha;\beta\gamma}_{\mr{Drude}}$, $ \sigma^{\alpha;\beta\gamma}_{\mr{BCD}}$, $\sigma^{\alpha;\beta\gamma}_{\mr{ChS}}$, and  $\sigma^{\alpha;\beta\gamma}_{\mr{gBC}}$ represent respectively the Matsubara term in nonlinear conductivity, the nonlinear Drude term, the Berry curvature dipole (BCD) term, the Christoffel symbol term, and the generalized Berry curvature term. Here, we newly elucidate last two terms, $\sigma^{\alpha;\beta\gamma}_{\mr{ChS}}$ and  $\sigma^{\alpha;\beta\gamma}_{\mr{gBC}}$, by considering the effect of dissipation. Under $\m{T}$-symmetry, $\sigma_{\mr{Drude}}$ and $\sigma_{\mr{ChS}}$ must be zero, while under $\m{PT}$-symmetry, $\sigma_{\mr{BCD}}$ and $\sigma_{\mr{gBC}}$ must be zero. We summarize $\tau$-dependence of the dominant contribution of each term in Table \ref{table:tau}. Below, we describe the detailed analysis of each term.

\begin{table}[h]
\renewcommand{\arraystretch}{1.2}
 \caption{Symmetry classification and $\tau$-dependence}
 \label{table:tau}
 \centering
 \scalebox{1.3}{
  \begin{tabular}{p{1.3cm}p{0.9cm}p{0.9cm}c}
   \hline
   term & $\m{T}$ & $\m{PT}$  & $\tau$-dependence \\
   \hline \hline
   $\sigma_{\mr{M}}$ & - & - & $\m{O}(\tau^{-1})$ \\
   $\sigma_{\mr{Drude}}$ & $\times$ & \checkmark & $\m{O}(\tau^{2})$ \\
   $\sigma_{\mr{BCD}}$ & \checkmark & $\times$ & $\m{O}(\tau)$ \\
   $\sigma_{\mr{ChS}}$ & $\times$ & \checkmark & $\m{O}(\tau^{0}) \ (\m{O}(\tau^{2}))$ \\
   $\sigma_{\mr{gBC}}$ & \checkmark & $\times$ & $\m{O}(\tau^{-1})$ \\
   \hline
  \end{tabular}}
  \renewcommand{\arraystretch}{1.0}
\end{table}

\subsection{analysis of each term}
In this subsection, we analyze each term in Eq.~(\ref{NLR}). Each term, except for the Matsubara term, is so complex, and therefore, we write the simple forms under some approximations, which are $\mr{Im}f(\epsilon_n+i\eta)\simeq \eta(\p f/\p\omega)_{\epsilon_n}$ and $\epsilon^2_{nm}\tau^2\gg1$. We write the detail derivation of each term and the full terms without approximation in appendix~\ref{app:NLR}. We also note that the approximation $\mr{Im}f(\epsilon_n+i\eta)\simeq \eta(\p f/\p\omega)_{\epsilon_n}$ can be justified even when $\beta\eta\sim0.5$.(see appendix~\ref{app:Linear}.)

\subsubsection{Matsubara term in nonlinear conductivity}
The Matsubara term in nonlinear coductivity is the contribution from the poles at Fermionic Matsubara frequencies as in the case of linear conductivity:
\begin{eqnarray}
   \sigma^{\alpha;\beta\gamma}_{\mr{M}} &=& \sum_{nml}\sum_{\omega_M>0}\mr{Re}\Bigl[\frac{\p}{\p\omega}\Bigl\{2\m{J}^{\alpha}_{nm}\frac{\p G^R_m}{\p\omega}\m{J}^{\beta}_{ml}G^R_l\m{J}^{\gamma}_{ln}(G^R_n\!-\!G^A_n)\nonumber\\
   &&+\m{J}^{\alpha}_{nm}\frac{\p G^R_m}{\p\omega}\m{J}^{\beta\gamma}_{mn}(G^R_n\!-\!G^A_n)\Bigr\} + \Bigl\{\beta\leftrightarrow\gamma\Bigr\}\Bigr]_{\omega = i\omega_M},\label{Matsubara2}
\end{eqnarray}
where $\m{J}^{\beta\gamma} = \p^{\beta\gamma}\m{H}_{\mr{eff}}$ and $\p^{\beta\gamma} = \p^{\beta}\p^{\gamma}$.
As we will numerically show, in the second-order conductivity under the time reversal symmetry and the condition $\pi k_BT>\eta$, the Matsubara term is small enough to be ignored, compared with the other finite terms, and therefore, the description by the Fermi DF of complex argument works well.

\subsubsection{nonlinear Drude term}
The nonlinear Drude term can be written as,
\begin{eqnarray}
   \sigma^{\alpha;\beta\gamma}_{\mr{Drude}}
   &\simeq& 2\sum_{\bk}\sum_n\tau^2\m{J}^{\alpha}_n \p^{\beta}\p^{\gamma} f(\epsilon_n).\label{Drude2_app}
\end{eqnarray}
Because the nonlinear Drude term is proportional to $\tau^2$, this term is most dominant in clean metals without TRS. We note that, if there is the band degeneracy at the Fermi surface, the Christoffel symbol term can be also dominant as we will show later.
\subsubsection{Berry curvature dipole term}
The BCD term\cite{PhysRevB.61.5337,PhysRevLett.115.216806} can be written as,
\begin{eqnarray}
   \sigma^{\alpha;\beta\gamma}_{\mr{BCD}} &=& \sigma^{\alpha;\beta\gamma}_{\mr{BCD:re}} + \sigma^{\alpha;\beta\gamma}_{\mr{BCD:im}}\label{BCD}\\ \sigma^{\alpha;\beta\gamma}_{\mr{BCD:re}}
   &\simeq& 2\tau\sum_{\bk}\sum_{nm}\p^{\gamma}(\Omega^{\alpha\beta}_{S;n,m})f(\epsilon_n)\label{BCDre_app}\\
   \sigma^{\alpha;\beta\gamma}_{\mr{BCD:im}}
   &\simeq&\sum_{\bk}\sum_{nm}
   \frac{\Omega^{\alpha\beta}_{S;n}\m{J}^{\gamma}_n}{\epsilon_{nm}\tau}\Bigl(\frac{\p^2 f}{\p\omega^2}\Bigr)_{\epsilon_{n}}\label{BCDim_app}\\
   \Omega^{\alpha\beta}_{S;n,m}&=& \frac{-i(\m{J}^{\alpha}_{nm}\m{J}^{\beta}_{mn}-\m{J}^{\beta}_{nm}\m{J}^{\alpha}_{mn})}{\epsilon^2_{nm}+4\eta^2}\label{sBC}
\end{eqnarray}
Under the TRS, the nonlinear Drude term must be zero and the BCD term is dominant. Because $\sigma^{\alpha;\beta\gamma}_{\mr{BCD:im}}$, which stems from the imaginary part of the Fermi DF, is proportional to $\eta$, it is not so large in clean systems. Here we can describe the BCD term by the smeared Berry curvature $\Omega^{\alpha\beta}_{S;n}$, which secures the convergence of the BCD term at the band-crossing points.
\subsubsection{Christoffel symbol term}
The Christoffel symbol term can be described as,
\begin{eqnarray}
   \sigma^{\alpha;\beta\gamma}_{\mr{ChS}}&=& \sigma^{\alpha;\beta\gamma}_{\mr{ChS:I}} + \sigma^{\alpha;\beta\gamma}_{\mr{ChS:I\hspace{-.01em}I}}\\
   \sigma^{\alpha;\beta\gamma}_{\mr{ChS:I}}
   &=& 2\sum_{\bk}\sum_n \Gamma^{\alpha;\beta\gamma}_{S;n} \Bigl(-\frac{\p f}{\p\omega}\Bigr)_{\epsilon_n} \label{ChS_app}\\
   \sigma^{\alpha;\beta\gamma}_{\mr{ChS:I\hspace{-.01em}I}}
   &\simeq&2\sum_{\bk}\sum_{n}\Gamma^{\alpha;\beta\gamma}_{S';n}\Bigl(-\frac{\p f}{\p\omega}\Bigr)_{\epsilon_n}\label{ChS2_app} 
\end{eqnarray}
where $\Gamma^{\alpha;\beta\gamma}_{S^{(')};n}$ is the smeared Christoffel symbol of the first kind\cite{PhysRevLett.112.166601,PhysRevX.10.041041}. Starting from the conventional Christoffel symbol $\Gamma^{\alpha;\beta\gamma}_{n}$, which reads,
\begin{eqnarray}
   \Gamma^{\alpha;\beta\gamma}_{n} &=& \frac{1}{2}\Bigl(\p^{\gamma}g^{\alpha\beta}_{n}+\p^{\beta}g^{\gamma\alpha}_{n}-\p^{\alpha}g^{\beta\gamma}_{n}\Bigr)\nonumber\\
   &=&\frac{1}{2}\Bigl(\braket{\p^{\alpha}n|\p^{\beta\gamma}n} +\braket{\p^{\beta\gamma}n|\p^{\alpha}n} \Bigr)\\
   &=& \sum_{m(\neq)n}\mr{Re}\Bigl[\frac{\m{J}^{\alpha}_{nm}}{\epsilon^2_{nm}}\Bigl(\sum_{l(\neq n)}\frac{\m{J}^{\beta}_{ml}\m{J}^{\gamma}_{ln} + (\beta\!\leftrightarrow\!\gamma)}{\epsilon_{nl}}\!+\!\m{J}^{\beta\gamma}_{mn}\Bigr)\Bigr],\nonumber\\
\end{eqnarray}
we define $\Gamma^{\alpha;\beta\gamma}_{S;n}$ by substituing $g^{\alpha\beta}_{n}\rightarrow g^{\alpha\beta}_{S;n}$, and $\Gamma^{\alpha;\beta\gamma}_{S';n}$ 
by substituting $1/\epsilon_{nl}\rightarrow \epsilon_{nl}/(\epsilon^2_{nl}+4\eta^2)$ and $1/\epsilon^2_{nm}\rightarrow 1/(\epsilon^2_{nm}+4\eta^2)$. $\sigma^{\alpha;\beta\gamma}_{\mr{ChS:I}}$ stems from the imaginary part of the Fermi FD in the term which is originally the BCD terms and the nonlinear Drude term, while $\sigma^{\alpha;\beta\gamma}_{\mr{ChS:I\hspace{-.01em}I}}$ stems from the full interband contribution.
Interestingly, in this regime, even though $\sigma^{\alpha;\beta\gamma}_{\mr{ChS:I}}$ stems from the imaginary part of the Fermi FD and the dissipation, it seems not to depend the dissipation strength. On the other hand, when we consider the band degeneracy at the Fermi surface $\epsilon_{n}=\epsilon_m\simeq0$, the Christoffel symbol term is proportional to $\tau^2$ because $g^{\alpha\beta}_{S;n}$ or $1/(\epsilon^2_{nm}+4\eta^2)$ is proportional to $\tau^2$, and therefore, the Christoffel symbol term also gives the multi-band correction to the nonlinear Drude term. We note that the difference between $\Gamma^{\alpha;\beta\gamma}_{S;n}$ and $\Gamma^{\alpha;\beta\gamma}_{S';n}$ appears when focusing on the band-degeneracy of the two-band systems with linear dispersion. In that case, $\Gamma^{\alpha;\beta\gamma}_{S;n}$ is finite and gives the correction to the nonlinear Drude term while $\Gamma^{\alpha;\beta\gamma}_{S';n}$ is zero.
We also note that this Christoffel symbol term is different from the one under magnetic field derived by the SCB treatment.\cite{PhysRevLett.112.166601}

\subsubsection{generalized Berry curvature term}
The generalized Berry curvature term can be written as,
\begin{eqnarray}
   \sigma^{\alpha;\beta\gamma}_{\mr{gBC}} = \sigma^{\alpha;\beta\gamma}_{\mr{gBC:re}} + \sigma^{\alpha;\beta\gamma}_{\mr{gBC:im}} + \sigma^{\alpha;\beta\gamma}_{\mr{gBC:add}}\label{gBC}
\end{eqnarray}
\begin{eqnarray}
   \sigma^{\alpha;\beta\gamma}_{\mr{gBC:re}}&\simeq&\sum_{\bk}\sum_{n,m(\neq n)}\frac{\Omega^{\alpha,\beta\gamma}_{S';n,m}}{\epsilon_{nm}\tau}\Bigl(-\frac{\p f}{\p\omega}\Bigr)_{\epsilon_n}\label{gBCD_app}\\
   \sigma^{\alpha;\beta\gamma}_{\mr{gBC:add}}&\simeq& 2\sum_{\bk}\sum_{n,m,l(\neq n)}\Bigl\{\frac{\mr{Im}(\m{J}^{\alpha}_{nm}\m{J}^{\beta}_{ml}\m{J}^{\gamma}_{ln})}{\epsilon^2_{nm}\epsilon_{nl}}\times\nonumber\\
   && \ \Bigl(\frac{1}{\epsilon_{nm}\tau}\!-\!\frac{1}{\epsilon_{nl}\tau}\Bigr)\Bigl(-\frac{\p f}{\p\omega}\Bigr)_{\epsilon_n} + (\beta\!\leftrightarrow\!\gamma)\Bigr\},\\
   \sigma^{\alpha;\beta\gamma}_{\mr{gBC:im}}
   &\simeq& -\sum_{\bk}\sum_{n}\frac{\Omega^{\alpha,\beta\gamma}_{S';n}}{\tau}\Bigl(-\frac{\p^2 f}{\p\omega^2}\Bigr)_{\epsilon_n},
\end{eqnarray}
 where $\Omega^{\alpha,\beta\gamma}_{S;n}$ is the smeared Berry curvature generalized to the second-order derivative and we define $\Omega^{\alpha;\beta\gamma}_{S';n,m}$ as derived from $\Omega^{\alpha;\beta\gamma}_{n,m}$, which reads,
 \begin{eqnarray}
    \Omega^{\alpha,\beta\gamma}_{n,m} &=& 2\mr{Im}\Bigl[\braket{\p^{\alpha}n|m}\braket{m|\p^{\beta\gamma}n}\Bigr]\\
   &=& \sum_{m(\neq)n}\mr{Im}\Bigl[\frac{\m{J}^{\alpha}_{nm}}{\epsilon^2_{nm}}\Bigl(\sum_{l(\neq n)}\frac{\m{J}^{\beta}_{ml}\m{J}^{\gamma}_{ln} + (\beta\!\leftrightarrow\!\gamma)}{\epsilon_{nl}}\!+\!\m{J}^{\beta\gamma}_{mn}\Bigr)\Bigr],\nonumber\\
   &&\label{gBC_f}
 \end{eqnarray}
 by substituting $1/\epsilon_{nl}\rightarrow \epsilon_{nl}/(\epsilon^2_{nl}+4\eta^2)$ and $1/\epsilon^2_{nm}\rightarrow 1/(\epsilon^2_{nm}+4\eta^2)$. $\sigma^{\alpha;\beta\gamma}_{\mr{gBC:add}}$ is zero when considering the two-band model and therefore it represent the more than two-band correction to $\sigma^{\alpha;\beta\gamma}_{\mr{gBC:re}}$.
 When we consider the nonreciprocal transport $\alpha=\beta=\gamma$ under the TRS, only this generalized Berry curvature term can be finite. In that case, the nonreciprocal conductivity is proportional to $\eta$ and the dissipation is essential for the nonreciprocal conductivity as pointed out in Ref.\cite{Morimoto2018}.
Moreover, we find that, when we focus on two-band models, higher order terms in momentum, such as $k^2$ term, in the Hamiltonian
is also necessary so that $J^{\alpha\alpha}_{mn} = \bra{m}(\partial^{\alpha\alpha}\m{H}_{eff})\ket{m}$ is nonzero,
for the diagonal part of the finite generalized Berry curvature with $\alpha=\beta=\gamma$.
It is because, $l$ is $m$ in Eq.~(\ref{gBC_f}) for the two-band model, and hence the diagonal part of the generalized Berry curvature vanishes as $\Omega^{\alpha;\alpha\alpha}_{n,m} = 2\Omega^{\alpha\alpha}_{n,m} \m{J}^{\alpha}/\epsilon_{nm} = 0$.\footnote{One can check $\Omega^{\alpha,\beta\gamma}_{n,m} = (\Omega^{\alpha\beta}_{n,m}\m{J}^{\gamma}_m+(\beta\leftrightarrow\gamma))/\epsilon_{nm}$ by substituting $l=m$ in Eq.~(\ref{gBC_f})}
 This means that the simple linear Weyl Hamiltonian cannot generate the nonreciprocal transport under the TRS. (see the detail in appendix~\ref{app:NLR}.)

\subsection{Model calculation}
\begin{figure}[t]
    \centering
    \includegraphics[width=0.96\linewidth]{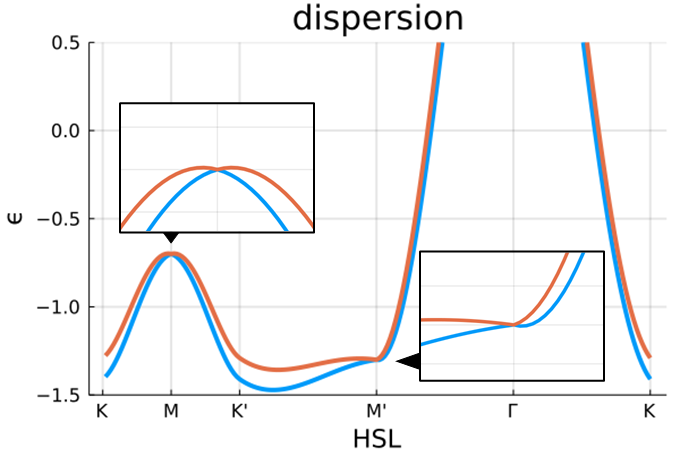}
    \caption{Dispersion of the model on the high symmetric line.\\
    we set $\mu=0$. The band degeneracy exists at $\epsilon=-0.7$, $\epsilon=-1.3$, and $\epsilon=2.7$ at $M$-, $M'$ and $\Gamma$-point.} 
    \label{fig:disp}
\end{figure}

Now we estimate how large the dissipation-induced geometric terms are in the model calculations.
We use the model which effectively describes 2D transition-metal dichalcogenides with uniaxial strain, such as MX$_2$ (M = Mo, W and X = S, Te)\cite{doi:10.1080/00018736900101307,C4CS00301B,PhysRevB.102.094507}, which reads,
\begin{eqnarray}
   \mathcal{H}_{eff} &=& \sum_{\bm{k},s,s'}
\Bigl(\left(\epsilon(\bm{k})\!-\!\mu\right)\sigma^0 + (\bm{h}\!+\!\bm{g}(\bm{k})) \cdot \bm{\sigma}\Bigr)_{ss'} c_{\bm{k},s}^{\dag} c_{\bm{k},s'}\label{app:TMD_NH}\\
\epsilon(\bk) &=& 2t\Bigl((1-p)\cos(\bk\cdot\bm{a_1})\nonumber\\
&& \ \ \ \ \ \ \ +\!\cos(\bk\cdot\bm{a_2})\!+\!\cos(\bk\cdot(\bm{a_1}\!+\!\bm{a_2}))\Bigr)\\
g^{x}(\bk) &=& \frac{\alpha_1}{2}\Bigl[\sin(\bk\cdot(\bm{a_1}+\bm{a_2})) + \sin(\bk\cdot\bm{a_2})\Bigr]\\
g^{y}(\bk) &=& -\frac{\alpha_1}{\sqrt{3}}\Bigl[ \sin(\bk\cdot\bm{a_1})\!+\!\frac{\sin(\bk\cdot(\bm{a_1}\!+\!\bm{a_2}))\!-\!\sin(\bk\cdot\bm{a_2})}{2}\Bigr]\nonumber\\
&&\\
g^z(\bk) &=& \frac{2\alpha_2}{3\sqrt{3}}\Bigl[\sin(\bk\cdot\bm{a_1})\!+\!\sin(\bk\cdot\bm{a_2})\!-\!\sin(\bk\cdot\bm{a_1}\!+\!\bm{a_2})\Bigr],\nonumber\\
\end{eqnarray}
where $\mu$ is the chemical potential, $\bm{h}$ is the magnetic filed, $t$ is the hopping, $\bm{a_1}=(1,0)$, $\bm{a_2}=(-0.5,\sqrt{3}/2)$, $p$ represents the effect of the uniaxial strain, and $\alpha_{1(2)}$ is the spin-orbit coupling. When $\bm{h}=0$, this model holds the time-reversal symmetry, while mirror symmetry of $y$-direction is broken due to finite $p$. In this section, we set the parameters as $t=0.5, p = 0.3, \alpha_1 = 0.08, \alpha_2=0.06$ in the numerical calculation for figure~\ref{fig:disp}, figure~\ref{fig:NLH}, \ref{fig:NRC}, and (\ref{fig:TRB_NRC}. Figure~\ref{fig:disp} shows the energy dispersion of the model when $\mu=0$. It has the band degeneracy $M$-, $M'$-, and $\Gamma$- point and their energy levels are $\epsilon=-0.7$, $\epsilon=-1.3$, $\epsilon=2.7$. The density of states is large near $\epsilon=-1.3$ corresponding to the nearly flat dispersion along $K'$-$M'$.
\subsubsection{Cases with the time-reversal symmetry}
\begin{figure}[t]
    \centering
    \includegraphics[width=0.96\linewidth]{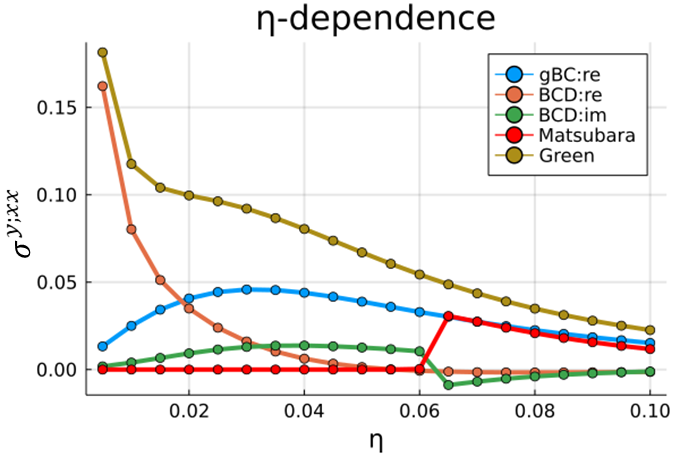}
    \includegraphics[width=0.96\linewidth]{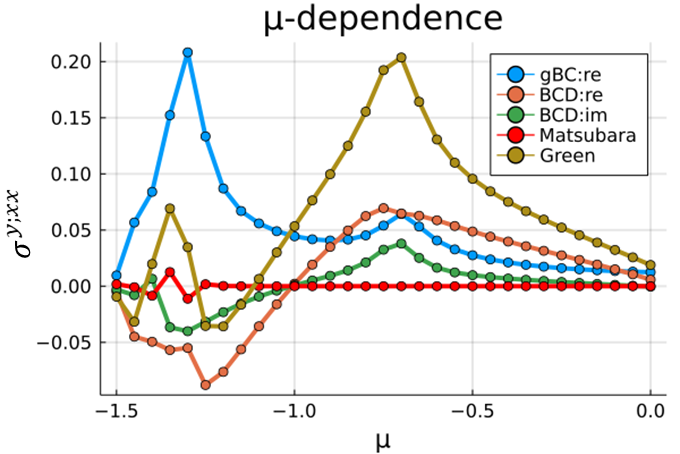}
    \caption{$\eta$-dependence and $\mu$-dependence of each contribution in nonlinear Hall conductivity.\\
    The top panel shows the dissipation-strength dependence of the nonlinear Hall conductivity $\sigma^{y;xx}$, and the bottom panel shows that the chemical potential dependence of $\sigma^{y;xx}$. We set $\mu=-0.9$ in the top panel and $\eta=0.02$ in the bottom panel. We perform the momentum integration by $1000\times1000$ and frequency integration by $1000$. The blue, orange, green, red, and brown plots respectively represent the generalized Berry curvature term, $\sigma_{\mr{BCD:re}}$, $\sigma_{\mr{BCD:im}}$, the Matsubara term, and the results by the Green function methods which coincides with the sum of all the  terms.} 
    \label{fig:NLH}
\end{figure}
In this subsection, we calculate the nonlinear Hall conductivity $\sigma^{y;xx}$ and the non-reciprocal conductivity $\sigma^{y;yy}$ in the model introduced above. We note that the model holds the mirror symmetry in $x$-direction and therefore $\sigma^{x;yy} = \sigma^{y;yx} =\sigma^{x;xx} =0$. In Figure.~\ref{fig:NLH}, we calculate $\eta$-dependence, and $\mu$-dependence of the nonlinear Hall conductivity $\sigma^{y;xx}$. Under the time-reversal symmetry, the dominant contribution are the BCD term in Eq.~(\ref{BCD})) and the generalized Berry curvature (gBC) term in Eq.~(\ref{gBC}). In the top panel of figure.\ref{fig:NLH}, as we have shown theoretically, $\sigma_{\mr{BCD:re}}$ is proportional to $1/\eta$, while $\sigma_{\mr{BCD:im}}$ and $\sigma_{\mr{gBC}}$ are proportional to $\eta$ in the regime $\eta<T$. In the limit $\eta \gg \epsilon_{nm}$, $\sigma_{\mr{BCD:im}}$ and $\sigma_{\mr{gBC}}$ are proportional to $1/\eta$, and therefore they both decrease in $\eta>0.04$. At $\eta = \pi T$, due to the singular behavior of $\mr{Im}(\p f/\p\omega)$, $\sigma_{\mr{BCD:im}}$ change its sign and the Matsubara term become large as to compensate it. 
The bottom panel of figure.\ref{fig:NLH} shows $\mu$-dependence. Around $\mu=-1.3$, there is quadratic dispersion at the Fermi surface around at $M$- and $M'$-point, and the gBC term becomes large. 
\begin{figure}[t]
    \centering
    \includegraphics[width=0.96\linewidth]{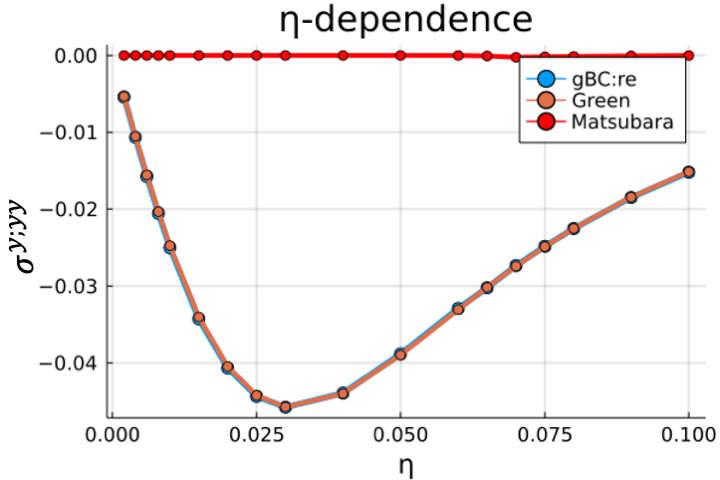}
    \includegraphics[width=0.96\linewidth]{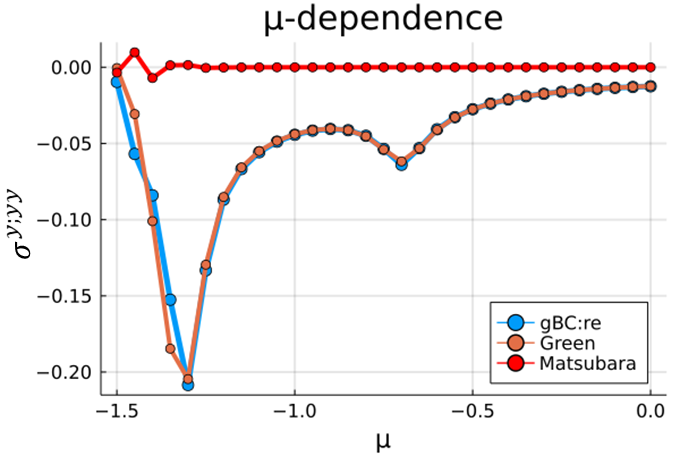}
    \caption{$\eta$-dependence and $\mu$-dependence of each contribution in nonreciprocal conductivity.\\
    The top panel shows the dissipation-strength dependence of the non-reciprocal conductivity $\sigma^{y;yy}$, and the bottom panel shows that the chemical potential dependence of $\sigma^{y;yy}$. We set $\mu=-0.9$ in the top panel and $\eta=0.02$ in the bottom panel. The blue, red, and orange plots respectively represent the generalized Berry curvature term, the Matsubara term, and the results by the Green function methods. We note that blue and orange plots overlap in the upper panel.} 
    \label{fig:NRC}
\end{figure}

Next, we calculate the non-reciprocal conductivity. As I have shown in the previous section, under the time-reversal symmetry, only the gBC term is finite. The top panel of figure.\ref{fig:NRC} shows the $\eta$-dependence of the non-reciprocal conductivity $\sigma^{y;yy}$. As in the case of the nonlinear Hall conductivity, the gBC term is proportional to $\eta$ for $\eta<T$ and proportional to $1/\eta$ for $\eta\gg\epsilon_{nm}$.

In the bottom panel of figure.\ref{fig:NRC}, the non-reciprocal conductivity (gBC term) behaves as same as the gBC term in the nonlinear Hall conductivity. (see the bottome panel of figure.\ref{fig:NLH}) 

\subsubsection{Cases without the time-reversal symmetry}
\begin{figure}[t]
    \centering
    \includegraphics[width=0.96\linewidth]{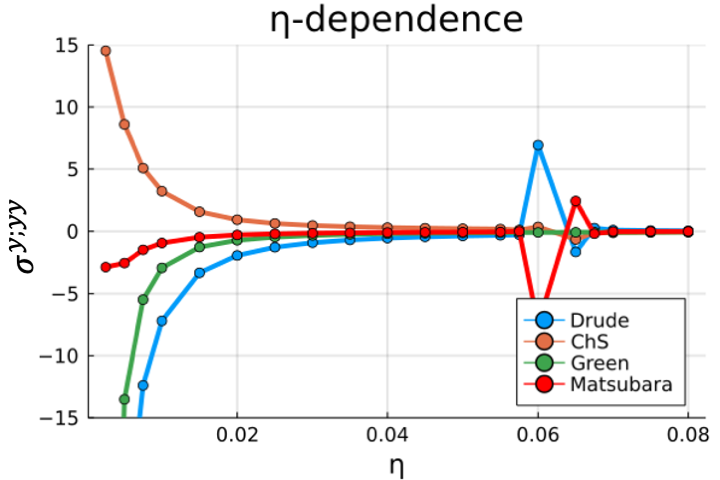}
    \includegraphics[width=0.96\linewidth]{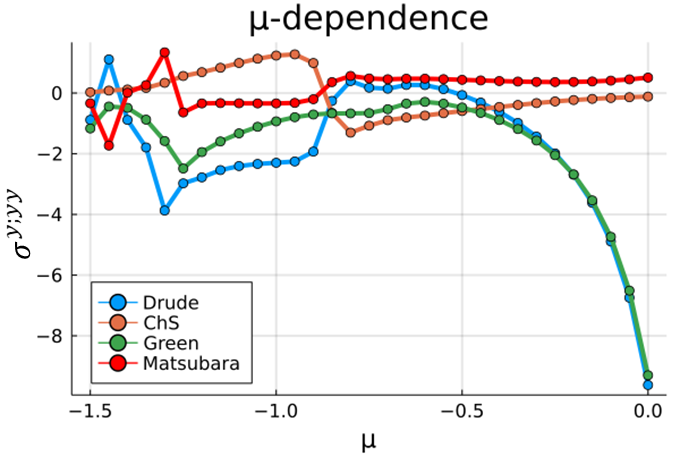}
    \caption{$\eta$-dependence and $\mu$-dependence of each contribution in nonreciprocal conductivity.\\
    The top panel shows the dissipation-strength dependence of the non-reciprocal conductivity $\sigma^{y;yy}$, and the bottom panel shows that the chemical potential dependence of $\sigma^{y;yy}$. We set $\mu=-0.9$ in the top panel and $\eta=0.02$ in the bottom panel. The blue, orange, red, and green plots respectively represent the Drude term, the Christoffel symbol term, the Matsubara term, and the results by the Green function methods.} 
    \label{fig:TRB_NRC}
\end{figure}
Next, we consider the case $\bm{h}\neq0$, in which time-reversal symmetry is broken. Without the time-reversal symmetry, the Drude term, the Christoffel symbol term can be finite. Here, we focus on the non-reciprocal conductivity with the magnetic field in $x$-direction $h_x = 0.05$ for figure~\ref{fig:TRB_NRC}.

In time-reversal symmetry broken systems, the (nonlinear) Drude term, which is proportional to $\tau^2$, is dominant in the small dissipation regime. (See the top panel of figure~\ref{fig:TRB_NRC}) The Christoffel symbol term seems also proportional to $\tau^2$ and gives the multi-band correction to the Drude term, because there is band degeneracy around at the Fermi surface.

The bottom panel of figure~\ref{fig:TRB_NRC} shows the $\mu$-dependence of the non-reciprocal conductivity, In the regime $\mu>-0.2$, the band velocity is large at the Fermi surface, and therefore the Drude term is dominant. On the other hand, in the regime $-0.85<\mu<-0.5$, the band velocity is not so large and there is the degeneration near the Fermi surface at $M'$-point, and therefore the Christoffel symbol term is dominant.

\section{nonlinear conductivity induced by dissipative quantum geometry in the Weyl Hamiltonian\label{Sec:Weyl}}
In this section, we derives the analytical results about the geometric terms derived in the previous sections.
\subsection{Hamiltonian and its dissipative quantum geometry}
Here, we consider the tilted Weyl Hamiltonian, which reads,
\begin{eqnarray}
   \m{H}(\bk) &=& (-\mu + \bm{t}\cdot\bk )\sigma^0 + t_0\bk\cdot\bm{\sigma},
\end{eqnarray}
where $\sigma^0$ is the two by two unit matrix, $\mu$ is the chemical potential, $t$ represents the tilting, $t_0$ is the Fermi velocity, $\bm{\sigma}=\{\sigma^x, \sigma^y, \sigma^z\}$ is the Pauli matrix. When $t<t_0$, this Hamiltonian describes the type-I Weyl Hamiltonian ,and when $t>t_0$, it describes the type-I\hspace{-.01em}I Weyl Hamiltonian. Here we set $\bm{t} = \{0,0,t\}$ for the simplicity.

In this Hamiltonian, the eigenvalues and the smeared geometrical quantities are,
\begin{eqnarray}
   E_{\pm} &=& tk_z\pm t_0k = t_0 k(\lambda_t\cos\theta \pm 1)\\
   \m{J}^{\alpha}_{\pm} &=& t_0 k^{\alpha} (\delta_{\alpha z}\lambda_t \pm1) \\
   \Omega^{\alpha\beta}_{S;+-} &=& -\Omega^{\alpha\beta}_{S;-+} = -\Omega^{\beta\alpha}_{S;+-} = \frac{k_{\gamma}\epsilon_{\alpha\beta\gamma}}{8(k^2 + \tilde{\eta}^2)k}\\
   g^{\alpha\beta}_{S;+-} &=& g^{\alpha\beta}_{S;-+} = g^{\beta\alpha}_{S;+-} = \frac{\delta_{\alpha\beta}(k^2-k^2_{\alpha})}{4(k^2 + \tilde{\eta}^2)k^2}
\end{eqnarray}
where $k = |k|$, $\lambda_t = t/t_0$, $\cos\theta = k_z/k$, and $\tilde{\eta} = \eta/t_0$ is the dissipation strength renormalized by the Fermi velocity.
We also define the geometric quantities, which appear in the nonlinear conductivity, as follows:
\begin{eqnarray}
   D^{\mu\nu}_{\pm} &\equiv& \frac{1}{2}\epsilon^{\mu\lambda\eta}\Omega^{\lambda\eta}_{S;\pm}\m{J}^{\nu}_{\pm} = \pm\frac{k_{\mu}(\delta_{z\nu}\lambda_t\pm k^{\nu}/k)}{8(k^2+\tilde{\eta}^2)k}\\
   F^{\mu\nu}_{\pm} &\equiv& \p^\mu g^{\nu\nu}_{S;\pm}\nonumber\\
   &=& -\frac{k^{\mu}}{2(k^2 + \tilde{\eta}^2)^2k^2}\Bigl\{(1-\frac{k^2_{\nu}}{k^2})(2k^2\!+\!\tilde{\eta}^2) + \delta_{\mu\nu}(k^2\!+\!\tilde{\eta}^2)\Bigr\}\nonumber\\
   &&\\
   \Gamma^{\mu;\nu\eta}_{S;\pm} &=& \frac{1}{2}(\delta_{\eta\mu}F^{\nu\eta}_{\pm} + \delta_{\mu\nu}F^{\eta\mu}_{\pm}-\delta_{\nu\eta}F^{\mu\nu}_{\pm})\\
   \Omega^{\alpha,\beta\gamma}_{S;\pm}&=& \frac{\pm k (\Omega^{\alpha\beta}_{S;\pm}\m{J}^{\gamma}_{\pm} + \Omega^{\alpha\gamma}_{S;\pm}\m{J}^{\beta}_{\pm})}{2(k^2+\tilde{\eta}^2)}
\end{eqnarray}
In the following, we suppose that $\beta\eta < 1$ and the broadening term can be ignored. We also approximate $\mr{Re}(-\p f/\p\omega)_{\epsilon_n+i\eta} \simeq (-\p f/\p\omega)_{\epsilon_n} \simeq \delta(\epsilon_n)$ and $\mr{Im}f(\epsilon_n+i\eta) \simeq \eta(\p f/\p\omega)_{\epsilon_n}$, and then, analyze the geometric terms in the type-I case ($0<\lambda_t<1$) and the type-I\hspace{-.01em}I case ($\lambda_t>1$).

\subsection{Type-I case}
We here consider the case where $\lambda_t<1$.
We can analytically calculate the BCD term and the Christffel symbol term in the limit $\eta\ll|\mu|$ and $\eta\gg|\mu|$ as,
\begin{align}
   &\sigma^{\alpha;\beta\gamma}_{\mr{BCD;I}}\nonumber\\ 
   &\simeq \tau\int \frac{\bm{dk}}{(2\pi)^3}  \Bigl\{\Omega^{\alpha\beta}_{S;\pm}\m{J}^{\gamma}_{\pm} + (\beta\leftrightarrow\gamma)\Bigr\}\delta(\epsilon_{\pm})\\
   &= \tau\int \frac{\bm{dk}}{(2\pi)^3}(\delta_{\gamma z}- \delta_{\beta z})\frac{1}{2}\epsilon_{\alpha\beta\gamma} (D^{zz}_{\pm})\\
   &=\frac{\tau}{2} \epsilon_{\alpha\beta\gamma} (\delta_{\gamma z}- \delta_{\beta z}) \times\label{BCD_app}\\
   & \ \ 
   \begin{dcases}
   \frac{1}{8\lambda_t}\Bigl\{2\lambda_t\!-\!\mr{ln}\Bigl(\frac{1\!+\!\lambda_t}{1\!-\!\lambda_t}\Bigr)\Bigr\} \ \ \ (\eta\ll|\mu|),\nonumber\\
    -\frac{1}{8\lambda_t^3}\frac{\mu^2}{\eta^2}\Bigl\{2\lambda_t\!-\!\tanh^{-1}\Bigl(\frac{2\lambda_t}{1\!+\!\lambda_t^2}\Bigr)\Bigr\} \ (\eta\gg|\mu|),
  \end{dcases}\nonumber
\end{align}
\begin{eqnarray}
   \sigma^{\alpha;\beta\gamma}_{\mr{ChS;I}}&\simeq&\int \frac{\bm{dk}}{(2\pi)^3}  \Gamma^{\alpha;\beta\gamma}_{S;\pm}\delta(\epsilon_{\pm})\\
   &=&\frac{1}{2}\Bigl(\delta_{\alpha\beta}\delta_{\gamma z} \tilde{F}^{\gamma\alpha}_{\mr{I}}\!+\!\delta_{\gamma\alpha}\delta_{\beta z} \tilde{F}^{\beta\gamma}_{\mr{I}}\!-\!\delta_{\beta\gamma}\delta_{\alpha z} \tilde{F}^{\alpha\beta}_{\mr{I}}\Bigr)\nonumber\\
\end{eqnarray}
with
\begin{eqnarray}
   && \tilde{F}^{zz}_{\mr{I}}\nonumber\\
   &&=
   \begin{dcases}
   \frac{\mu}{t_0\lambda_t}\Bigl\{-\frac{6}{\lambda_t^2} +\frac{3(2-\lambda_t^2)}{2\lambda_t^3} \mr{ln}\Bigl(\frac{1+\lambda_t}{1-\lambda_t}\Bigr)+\frac{1}{1-\lambda_t^2}\Bigr\}\label{ChS0_app}\\
    \ \ \ \ \ \ \ \ \ \ \ \ \ \ \ \ \ \ \ \ \ \ \ \ \ \ \ \ \ \ \ \ \ \ \ \ \ \ \ \ \ \ \ \ \ \ \ \ (\eta\ll|\mu|),\\
   \frac{(3-\lambda_t^2)\mu}{t_0\lambda_t^3}\Bigl\{\frac{1}{2\lambda_t}\tanh^{-1}\Bigl(\frac{2\lambda_t}{1+\lambda_t^2}\Bigr) - \frac{1}{1-\lambda_t^2}\Bigr\} \\             
    \ \ \ \ \ \ \ \ \ \ \ \ \ \ \ \ \ \ \ \ \ \ \ \ \ \ \ \ \ \ \ \ \ \ \ \ \ \ \ \ \ \ \ \ \ \ \ \  (\eta\gg|\mu|),
   \end{dcases}\\
   && \tilde{F}^{zx}_{\mr{I}}=\tilde{F}^{zy}_{\mr{I}}=\tilde{F}^{xz}_{\mr{I}}=\tilde{F}^{yz}_{\mr{I}}\nonumber\\
   &&=
   \begin{dcases}
   \frac{\mu}{t_0\lambda_t}\Bigl\{-\frac{6}{\lambda_t^2} +\frac{3(2-\lambda_t^2)}{2\lambda_t^3} \mr{ln}\Bigl(\frac{1+\lambda_t}{1-\lambda_t}\Bigr)+\frac{1}{1-\lambda_t^2}\Bigr\}\label{ChS0x_app}\\
    \ \ \ \ \ \ \ \ \ \ \ \ \ \ \ \ \ \ \ \ \ \ \ \ \ \ \ \ \ \ \ \ \ \ \ \ \ \ \ \ \ \ \ \ \ \ \ \  (\eta\ll|\mu|), \\
   \frac{(3-\lambda_t^2)\mu}{t_0\lambda_t^3}\Bigl\{\frac{1}{2\lambda_t}\tanh^{-1}\Bigl(\frac{2\lambda_t}{1+\lambda_t^2}\Bigr) - \frac{1}{1-\lambda_t^2}\Bigr\} \\
    \ \ \ \ \ \ \ \ \ \ \ \ \ \ \ \ \ \ \ \ \ \ \ \ \ \ \ \ \ \ \ \ \ \ \ \ \ \ \ \ \ \ \ \ \ \ \ \  (\eta\gg|\mu|),
   \end{dcases}
\end{eqnarray}
Here we omit the analysis of the gBC term, because here we approximate $\mr{Re}(-\p f/\p\omega)_{\epsilon_n+i\eta}\simeq\delta(\epsilon)$. This approximation is justified when $T\rightarrow0$ and $\eta\rightarrow0$ with $k_BT\gg\eta$, while the gBC term is proportional to $\eta$.
Interestingly, the nonlinear Hall conductivity by the BCD term is independent of the chemical potential in the regime $|\mu|\gg\eta$, while it is proportional to $\mu^2/\eta^2$ in the regime $|\mu|\ll\eta$, which results in the dip around $|\mu|<\eta$. 
The Christoffel symbol term is proportional to $\mu$ in both regime. 
We note that, although Eq.~(\ref{BCD_app}) and Eq.~(\ref{ChS0_app}) appears to diverge at $\lambda_t\rightarrow0$, they becomes zero by appropriately expanding the terms in $\{\dots\}$ in Eq.~(\ref{BCD_app}) and Eq.~(\ref{ChS0_app}).

\subsection{Type-I\hspace{-.01em}I case}
Next, we analyze the type-I\hspace{-.01em}I Weyl Hamiltonian where $\lambda_t>1$. Although the only difference in the model from type-I is the magnitude of $\lambda_t$, the behavior is completely different. Because the integral of momentum space $\int_0^{\infty}dk k^2$ is not convergent. Therefore, we introduce the cut-off scale $\Lambda$ ($\int_0^{\infty}dk\rightarrow\int_0^{\Lambda}dk$), in which the approximation to the Weyl Hamiltonian is justified. Then, we can derive the BCD term and the Christoffel symbol term as,
\begin{eqnarray}
   \sigma^{\alpha;\beta\gamma}_{\mr{BCD;I\hspace{-.01em}I}}&=&-\frac{\tau\epsilon_{\alpha\beta\gamma} (\delta_{\gamma z}- \delta_{\beta z})}{32\pi^2\lambda_t}\mr{ln}\Bigl(\frac{\Lambda^2+\eta^2}{(\mu/(1+\lambda_t))^2 +\eta^2}\Bigr)\nonumber\\
   &&\\
   \sigma^{\alpha;\beta\gamma}_{\mr{ChS;I\hspace{-.01em}I}}
   &=&\frac{1}{2}\Bigl(\delta_{\alpha\beta}\delta_{\gamma z} \tilde{F}^{\gamma\alpha}_{\mr{I\hspace{-.01em}I}}\!+\!\delta_{\gamma\alpha}\delta_{\beta z} \tilde{F}^{\beta\gamma}_{\mr{I\hspace{-.01em}I}}\!-\!\delta_{\beta\gamma}\delta_{\alpha z} \tilde{F}^{\alpha\beta}_{\mr{I\hspace{-.01em}I}}\Bigr)\\
   \tilde{F}^{zz}_{\mr{I\hspace{-.01em}I}}&=& \frac{\Lambda}{t_0}\Bigl(\frac{3}{2\lambda_t^2}-\frac{1}{\lambda_t^4}\Bigr) + \m{O}(\mr{ln}\Lambda)\\
   \tilde{F}^{zx}_{\mr{I\hspace{-.01em}I}}&=&\tilde{F}^{zy}_{\mr{I\hspace{-.01em}I}}=\tilde{F}^{xz}_{\mr{I\hspace{-.01em}I}}=\tilde{F}^{yz}_{\mr{I\hspace{-.01em}I}}= \frac{\Lambda}{t_0}\Bigl(\frac{1}{\lambda_t^2}-\frac{1}{\lambda_t^4}\Bigr) + \m{O}(\mr{ln}\Lambda)\nonumber\\
\end{eqnarray}
In the type-I\hspace{-.01em}I case, when we consider the limit $\mu\rightarrow0$ with $\eta/\mu= \mr{const.}$, the BCD term shows a logarithmic divergence, while it is convergence with finite $\eta$. This behavior is completely different from the type-I case, where the BCD term has the dip at $\mu=0$. In the limit $\Lambda\gg t_0,\mu,\eta,\lambda_t$, the dominant term of the Christoffel symbol term is independent of $\mu$ and $\eta$ in the type-I\hspace{-.01em}I Weyl Hamiltonian.

\begin{figure}[t]
    \centering
    \includegraphics[width=0.96\linewidth]{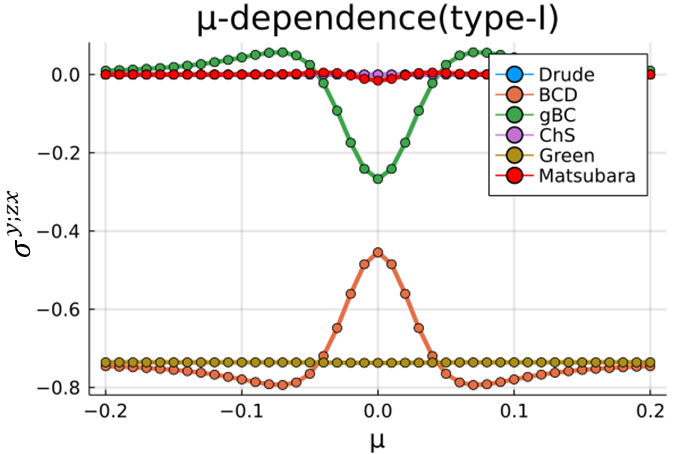}
    \includegraphics[width=0.96\linewidth]{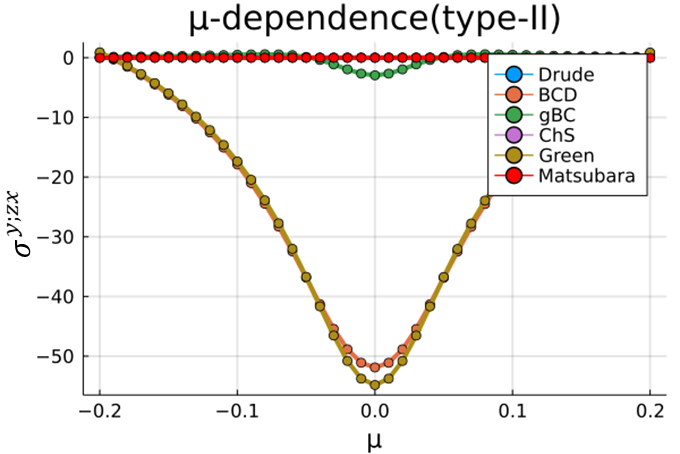}
    \caption{$\mu$-dependence of the nonlinear Hall conductivity in type-I and I\hspace{-.01em}I Weyl Hamiltonian.\\
    The top (bottom) panel shows the nonlinear Hall conductivity in the type-I (I\hspace{-.01em}I) Weyl Hamiltonian. We set the parameter as $t_0=1.0, t=0.3$ for the type-I Weyl Hamiltonian and $t_0=0.3, t=0.5$ for the type-I\hspace{-.01em}I Weyl Hamiltonian. We also set $k_BT=0.02$ and $\eta=0.01$ for each case. The terms except the BCD term and the gBC term are almost zero.} 
    \label{fig:NLHinWeyl}
\end{figure}

\subsection{Numerical results}
We also check the numerical calculation about the chemical potential dependence of the nonlinear Hall effect $\sigma^{y;zx} = -\sigma^{x;yz}$.
Interestingly, in addition to the $\mu$-independence of the nonlinear Hall conductivity at $|\mu|\gg\eta$ as we have analytically shown, the gBC term compensates the dip of the BCD term around $\mu=0$, which results in complete $\mu$-independence of the nonlinear Hall conductivity.(See the top panel of figure~\ref{fig:NLHinWeyl}.)
The bottom panel of figure~\ref{fig:NLHinWeyl} shows that the peak behaviour of the nonlinear Hall conductivity at $\mu=0$ as we have shown, and its order of the conductivity is much larger than type-I Weyl systems. This result means that the type-I\hspace{-.01em}I Weyl materials can show large nonlinear Hall effect.  

We again stress that this chemical potential independence or the peak behavior of the nonlinear Hall conductivity can be highly utilized for the detection of the Weyl points and their type.

\section{Outlook and Summary\label{Sec:Summary}}
In this paper, we have analyzed the dissipation effect on the linear and nonlinear DC conductivity under the Markov approximation in multi-band systems. 

Starting from the Green function formalism, we elucidates the effect of the dissipation: the shift of the distribution function in the imaginary direction and the Matsubara term, which cannot be included by the conventional methods such as the SCB treatment and the reduced density matrix methods. Moreover, we clarify that the novel terms from the imaginary part of the distribution function also have the geometric nature, as the quantum metric term in the linear conductivity, and the Christoffel symbol term and the generalized Berry curvature term in the nonlinear conductivity. These terms give the multi-band correction to the (nonlinear) Drude term when there is the band degeneracy at the Fermi surface. Although the Matsubara term is not small in the linear response, it is small enough to be ignored in nonlinear response especially under the TRS, and therefore, the description of the Fermi DF of complex argument works well. Under the TRS, the inversion symmetry breaking is encoded in the multi-band effect such as the BCD and the generalized Berry curvature. At the pole of the Matsubara frequency, the large imaginary value of $i\omega_M$ in denominator of the $G(i\omega_M)$ cloaks such kind of the geometric structure, and therefore, the Matsubara term becomes almost zero. When $\pi k_BT\sim \eta$, $i\omega_M-i\eta$ in the denominator of $G^A(i\omega_M)$ becomes almost zero and the above cloaking is unveiled, and the Matsubara term becomes finite. In the nonlinear conductivity without TRS and in the linear conductivity, the Drude term, which is the intra-band contribution and does not need the geometric structure, can be finite, and therefore, the Matsubara term can be also finite.

We have also elucidated the geometric origin of the non-reciprocal conductivity under the TRS, which can be described by the Berry curvature generalized to the higher-order derivative. For two-band systems, we have identified another condition of the non-reciprocal conductivity under the time-reversal symmetry, that is the quadratic term, in addition to the dissipation which is pointed out in Ref.\cite{Morimoto2018}.

Then, we numerically calculate the $\eta$- and $\mu$-dependence of each contribution in the model which describes TMD materials. The results shows that the Matsubara term becomes large at $\eta > \pi k_BT$ when the terms from the imaginary part of the distribution function is not small. We have shown that, in some regime, the novel geometric term become dominant.

Finally, we analyze each geometric terms in the Weyl Hamiltonian. We clarify that the chemical potential dependence around at the Weyl point of nonlinear Hall conductivity is completely different between the type-I and type-I\hspace{-.01em}I case. While the nonlinear Hall conductivity is comletely independent of teh chemical potential in the type-I case, it shows the logarithmic divergence behavior at the energy level where the Weyl points exist in the type-I\hspace{-.01em}I case. This result suggest that the detection of the chemical potential dependence of the nonlinear Hall conductivity under the time-reversal symmetry can be utilized to detect the existence of the Weyl points and their type. Moreover, we also show that the type-I\hspace{-.01em}I Weyl materials can show large nonolinear Hall conductivity due to its divergent behaviour at the Weyl points.

Although we consider the DC conductivity in this paper, the similar analysis can be applied to the photovoltaic effect, in which the dissipation holds the important role. It is left for the future works. 

\section*{acknowledge}
YM deeply appreciate to Hikaru Watanabe, Yoichi Yanase, and Kazuaki Takasan for fruitful discussions. YM also thanks Masahiko G. Yamada, Atsuo Shitade, and Hiroshi Shinaoka for their advice about the numerical calculations. We performed all calculation with Julia\cite{bezanson2017julia}. The code which we used is uploaded to https://github.com/YoshihiroMichishita/julia. This work is supported by the WISE program, MEXT. YM is supported by a JSPS research fellowship and by JSPS KAKENHI (Grant No. 20J12265).  Naoto Nagaosa is supported by JSPS KAKENHI Grant
(No. 18H03676), and by JST CREST Grant Number JPMJCR1874, Japan.

\appendix
\section{Detail derivation of the dissipation-induced geometric term in linear conductivity\label{app:Linear}}
Contribution from the pole of the advanced Green functions to the linear conductivity can be described as follows, 
    \begin{eqnarray}
       \sigma^{\alpha\beta}_{\mr{G}}&=& \sigma^{\alpha\beta}_{\mr{Drude}} + \sigma^{\alpha\beta}_{\mr{QM:re}} +\sigma^{\alpha\beta}_{\mr{QM:im}}\\
       \sigma^{\alpha\beta}_{\mr{Drude}}&=&\sum_{\bk}\sum_n\m{J}^{\alpha}_n\m{J}^{\beta}_n\tau \mr{Re}\Bigl(-\frac{\p f}{\p\omega}\Bigr)_{\epsilon_n+i\eta}\label{L_BI_Dr}\\
       \sigma^{\alpha\beta}_{\mr{QM:re}}&=&\sum_{\bk}\sum_n\frac{g^{\alpha\beta}_{S;n}}{\tau} \mr{Re}\Bigl(-\frac{\p f}{\p\omega}\Bigr)_{\epsilon_n+i\eta}\label{app:L_BI_QMre}\\
       \sigma^{\alpha\beta}_{\mr{QM:im}}&=&\sum_{\bk}\sum_n\sum_{m\neq n}g^{\alpha\beta}_{S;n,m}\epsilon_{nm}\mr{Im}\Bigl(-\frac{\p f}{\p\omega}\Bigr)|_{\epsilon_n+i\eta}\label{BI_linear}\\
        g^{\alpha\beta}_{S;n,m}&=& \frac{(\m{J}^{\alpha}_{nm}\m{J}^{\beta}_{mn}+\m{J}^{\beta}_{nm}\m{J}^{\alpha}_{mn})}{2(\epsilon^2_{nm} + 4\eta^2)}. \label{sQ}
    \end{eqnarray}
We call $g^{\alpha\beta}_{S;n}$ as ``the smeared quantum metric.'' We note that the smeared geometric quantities are well defined and finite at gapless points where $\epsilon_{nm}=0$. 
$\sigma^{\alpha\beta}_{\mr{QM:re}} +\sigma^{\alpha\beta}_{\mr{QM:im}}$ is the novel term, which we call ``quantum metric term'' at the Fermi surface.  

\begin{figure}[t]
    \centering
    \includegraphics[width=0.49\linewidth]{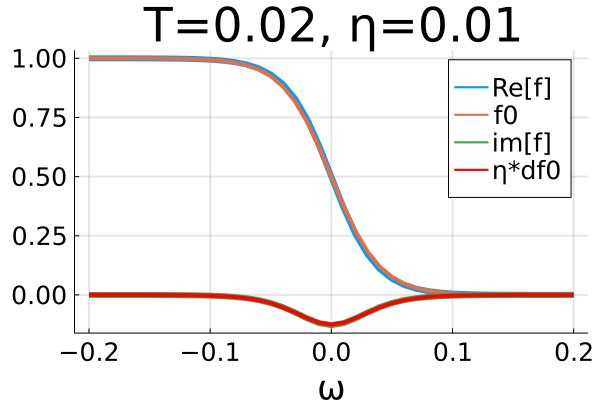}
    \includegraphics[width=0.49\linewidth]{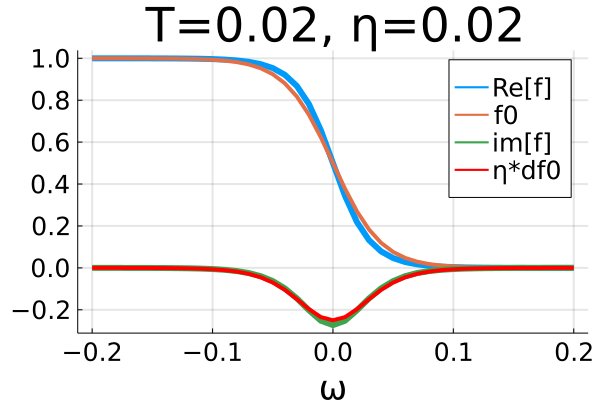}
    \caption{Approximation about the imaginary part of the distribution function.\\
    The blue(orange) plots show the real(imaginary) part of $f(\omega+i\eta)$. The green plots show $f(\omega)$ and the red plots show $\eta\p f/\p\omega$.  The parameters are $T = 0.02$, and $\eta=0.01$ on the left panel and $\eta=0.02$ on the right panel. This figure shows that the approximation $\mr{Re}f(\omega+i\eta)\simeq f(\omega)$ and $\mr{Im}f(\omega+i\eta)\simeq \eta\p f(\omega)/\p\omega$ is enough good even when $\beta\eta\sim 0.5$.} 
    \label{fig:Im_FD}
\end{figure}

We note that we can approximate $\mr{Re}f(\epsilon_n+i\eta)\simeq f(\epsilon_n)$ and $\mr{Im}f(\epsilon_n+i\eta)\simeq \eta \p f(\epsilon_n)/\p\omega$ when $\beta\eta \ll 1$. Figure.~\ref{fig:Im_FD} shows that this approximation is valid even when $\beta\eta\sim0.5$. Under this approximation, in the both limits $\epsilon_{nm}\gg \eta$ and $\epsilon_{nm}\rightarrow0$, the terms proportional to the imaginary part of the distribution function can be written as,
\begin{align}
   &g^{\alpha\beta}_{S;n,m}\epsilon_{nm}\mr{Im}\Bigl(-\frac{\p f}{\p\omega}\Bigr)|_{\epsilon_n\!+\!i\eta} \simeq \frac{\epsilon_{nm}}{2\tau}g^{\alpha\beta}_{S;n,m}\Bigl(-\frac{\p^2 f}{\p\omega^2}\Bigr)|_{\epsilon_n}
\end{align}
When we see the frequency derivative of the Fermi distribution function as $\sim\m{O}(\beta)$, the imaginary part contributions are the order $\sim\m{O}(\epsilon_{nm}\beta)$ compared to the real part of the contribution. This means that the contribution from the imaginary part of the distribution function can be dominant at low temperature. We note that, under the approximation $\mr{Re}f(\epsilon_n+i\eta)\simeq f(\epsilon_n)$ and $\mr{Im}f(\epsilon_n+i\eta)\simeq \eta \p f(\epsilon_n)/\p\omega$, Eq.~(\ref{app:L_BI_QMre}) can be derived from the RDM methods by substituting $f(\epsilon_n)\rightarrow f(\epsilon_n+i\eta)$ in the anomalous quantum Hall term and considering the imaginary part of the Fermi distribution function. From this point of view, $\sigma^{\alpha\beta}_{\mr{QM:re}}$ also originates from the imaginary part of the distribution function in the anomalous quantum Hall term.

When considering the band degeneracy at the Fermi surface which means $\epsilon_n=\epsilon_m$, the smeared quantum metric $g^{\alpha\beta}_{S;n}$ is proportional to  $\tau^2$ and the qunatum metric term in Eq.~(\ref{L_BI_QMre}) is proportional to $\tau$, which means the quantum metric term gives the multi-band correction to the Drude term in Eq.~(\ref{L_BI_Dr}). Therefore, the quantum metric term becomes important in the materials which have the large band degeneracy at the Fermi surface or the strongly-dissipative system.

\section{Intuitive understanding about the broadening term\label{app:Intuitive}}
\begin{figure}[t]
    \centering
    \includegraphics[width=0.48\linewidth]{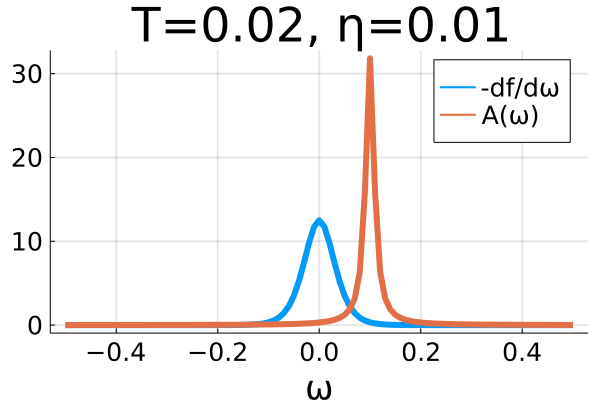}
    \includegraphics[width=0.48\linewidth]{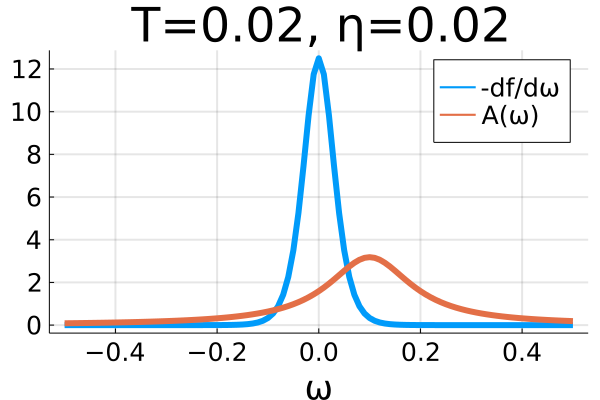}
    \caption{broadening of the spectral function v.s. derivative of the Fermi DF.\\
    The blue(orange) plots show the $\omega$-derivative of the Fermi FD $(-\p f/\p\omega)$ (the spectral function $A(\omega)$). Here we set the temperature $T=0.02$, and $\epsilon_n = 0.1$. $\eta = 0.01 (0.1)$ in the left(right) panel.} 
    \label{fig:SoDOS}
\end{figure}

In this part, we give the intuitive understanding about the shift of the Fermi DF and the broadening term. In the conventional band representation, the occupancy of the band is determined by its energy and the Fermi DF ($f(\epsilon_n)$), and it is true in the limit of $\eta\rightarrow0$. However, when $\eta$ is finite, there is the broadening of the spectral function and the occupancy of the band cannot be determined just by the band energy and the Fermi DF. In the small dissipation regime $\eta\ll T$, the broadening is small and the bands, who have the energy level where $(df/d\omega)_{\epsilon_n}$ is finite, contribute to the transport. For an example, in Figure.~(\ref{fig:SoDOS}), the band, whose energy is $|\epsilon_n|>T$, almost does not contribute the transport. In this case, the description by the Fermi DF is accurate, and the shift of the Fermi DF to the imaginary direction well approximates the broadening of the spectral function. On the other hand, when we consider the highly dissipative case $\eta > T$, even though the band energy is far away from the Fermi energy $|\epsilon_n|\gg T$ and $(df/d\omega)_{\epsilon_n}\simeq0$, there is the overlap between $df(\omega)/d\omega$ and $A(\omega)$, and this band can contribute to the transport.
In this case, the description by the Fermi DF becomes bad, and the Matsubara term, which is not described by the Fermi DF, can become large. These understanding should also hold true in nonlinear conductivity. 

\section{model calculation\label{app:linear_num}}
We numerically calculate these contribution to the linear conductivity in the nodal-line semimetals, which is described by the following Hamiltonian as\cite{PhysRevResearch.2.043311},
\begin{eqnarray}
   \m{H} &=& \mu\tau^0 + t(2+\cos{k_0}-\cos{k_x}-\cos{k_y}-\cos{k_z})\tau^z\nonumber\\
   && \ \ + v\sin{k_z}\tau^y + \Delta\tau^x. \label{Ham_NLSM}
\end{eqnarray}
\begin{figure}[t]
    \centering
    \includegraphics[width=0.96\linewidth]{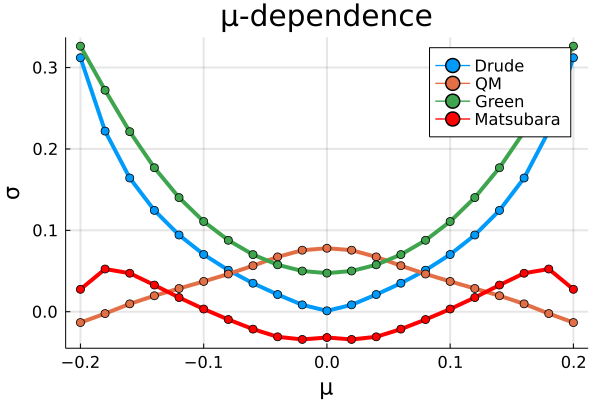}
    \includegraphics[width=0.96\linewidth]{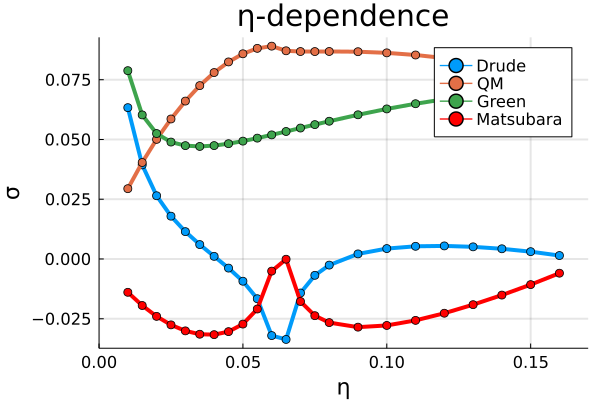}
    \caption{Linear conductivity in a nodal-line semimetal.\\
    The parameters are $T = 0.02, \eta=0.04$ on the top panel and $T=0.02, \mu=0$ on the bottom panel. We perform the momentum integration by $400\times400\times400$ and frequency integration by $1000$. The blue, orange, green, red plots respectively describe the Drude term, the quantum metric term ($\sigma^{zz}_{\mr{QM:re}}+\sigma^{zz}_{\mr{QM:im}}$), the total conductivity calculated by the Green function methods, and the Matsubara term.} 
    \label{fig:mudep}
\end{figure}
$\mu, t, v, \Delta$ represent the chemical potential, the hopping amplitude, the hybridization, and the TRS breaking parameter. 
We note that this tight-binding Hamiltonian
describes the low-energy dispersion near the Fermi level of
CaAgP and Ca$_3$P$_2$\cite{doi:10.7566/JPSJ.85.013708,PhysRevB.93.205132}. For the upper panel of figure~\ref{fig:mudep}, the chemical potential ($\mu$) dependence of the linear conductivity is shown. Here the temperature $k_BT= 0.02$, $\eta=0.04$, and $\Delta=0$.  
This model has the nodal line at the Fermi surface when $\mu=0, \Delta=0$ and the quantum metric term becomes large. We note that, in this model, $\m{J}^z_{nn}$ is zero on the nodal line and therefore the Drude term becomes almost zero at $\mu=0$.

In this case, the quantum metric term is dominant at $\mu=0$ and the Matsubara term is less than 0 to
compensate the overestimated quantum metric term in the description of the Fermi distribution function. Even when the system has a quadratic band-touching at the Fermi surface and the Drude term is zero, if these band are constructed by the hybridization, conductivity can be finite due to the quantum metric term. 

The bottom panel of figure~\ref{fig:mudep} shows that $\eta$-dependence of the linear conductivity. Here the temperature $T= 0.02$, $\mu=0$, and $\Delta=0$. In the small dissipation regime, the quantum metric term and the Matsubara term is proportional to $\eta$, and therefore, the Drude term is dominant. In our formulation, $\eta = \pi T$, where the Fermi distribution function $f(\epsilon_n + i\eta)$ behaves as the Bose distribution function $f_B(\epsilon_n)$, is a singular. In the large dissipation regime where $\eta > \pi T$, contribution from the Drude term becomes almost zero. This large dissipation regime at low temperature can be realized in the quantum critical regime, because the large quantum fluctuation behaves as the large dissipation from the single-particle points of view. In such regime, the SCB treatment and also the reduced density matrix method is not appropriate. Therefore, we must use the Green function methods in such regime.

The relaxation time $\tau$ in usual materials is about $1\sim100$ picosecond\cite{Du2019}, which equals to $\eta\simeq 2.1 * 10^{-5}\sim 2.1*10^{-3}$[eV]. In this case, the temperature which satisfies $\pi k_B T > \eta$ is about $T > 7.5*10^{-2}\sim 7.5$[K], and this condition is often not satisfied in the condensed matter physics. Therefore, our result shows that we should consider the broadening of the spectral function or to say the quantum dissipation at low temperature, and the SCB treatment or the reduced density matrix formalism is not appropriate especially when we consider the transport at a degenerated Fermi surface.

\section{Detail derivation of the dissipation-induced geometric terms in nonlinear conductivity\label{app:NLR}}
The nonlinear conductivity is described with the Green function formalism as
\begin{eqnarray}
   &&\sigma^{\alpha;\beta\gamma}_{\mr{DC}}\nonumber\\
   &&= \int_{-\infty}^{\infty} \frac{d\omega}{2\pi}\mr{Im}\mr{Tr}\Bigl[\Bigl\{2\m{J}^{\alpha}_{nm}\frac{\p G^R_m}{\p\omega}\m{J}^{\beta}_{ml}G^R_l\m{J}^{\gamma}_{ln}(G^R_n-G^A_n)\nonumber\\
   && \ \ \ \ \ \ \ \ \ \ + \m{J}^{\alpha}_{nm}\frac{\p G^R_m}{\p\omega}\m{J}^{\beta\gamma}_{mn}(G^R_n-G^A_n)\Bigr\} + \{\beta\leftrightarrow\gamma\}\Bigr],
\end{eqnarray}
where $\m{J}^{\beta\gamma} = \p^{\beta}\p^{\gamma}\m{H}_{\mr{eff}}$. By starting from this Green function formalism and performing the same procedure as in the linear case, we can derive the following equation as,
\begin{widetext}
    \begin{eqnarray}
       \sigma^{\alpha;\beta\gamma}_{\mr{DC}}&=& \sum_{\bk}\Bigl( \sigma^{\alpha;\beta\gamma}_{\mr{M}} + \tilde{\sigma}^{\alpha;\beta\gamma}_{\mr{Drude}} + \tilde{\sigma}^{\alpha;\beta\gamma}_{\mr{BCD}} + + 
       \tilde{\sigma}^{\alpha;\beta\gamma}_{\mr{Inter}}\Bigr),\label{second}\\
       \sigma^{\alpha;\beta\gamma}_{\mr{Th}}&=& \sum_{nml}\sum_{\omega_M>0}\mr{Re}\Bigl[\frac{\p}{\p\omega}\Bigl\{2\m{J}^{\alpha}_{nm}\frac{\p G^R_m}{\p\omega}\m{J}^{\beta}_{ml}G^R_l\m{J}^{\gamma}_{ln}(G^R_n-G^A_n)+\m{J}^{\alpha}_{nm}\frac{\p G^R_m}{\p\omega}\m{J}^{\beta\gamma}_{mn}(G^R_n-G^A_n)\Bigr\} + \bigl\{\beta\leftrightarrow\gamma\Bigr\}\Bigr]_{\omega = i\omega_M}\label{thermal2}\\
       \tilde{\sigma}^{\alpha;\beta\gamma}_{\mr{Drude}}&=&2 \sum_{n}\tau^2\mr{Re}\Bigl[\m{J}^{\alpha}_n\m{J}^{\beta}_n\m{J}^{\gamma}_n\Bigl(-\frac{\p^2f}{\p\omega^2}\Bigr) + \m{J}^{\alpha}_n\m{J}^{\beta\gamma}_n\Bigl(-\frac{\p f}{\p\omega}\Bigr) + \sum_{m(\neq n)}\m{J}^{\alpha}_n\frac{\m{J}^{\beta}_{nm}\m{J}^{\gamma}_{mn} + (\beta\leftrightarrow\gamma)}{\epsilon_{nm} + 2i\eta}\Bigl(-\frac{\p f}{\p\omega}\Bigr)\Bigr]_{\omega=\epsilon_n+i\eta}\label{Drude2}\\
       \tilde{\sigma}^{\alpha;\beta\gamma}_{\mr{BCD}}&=& 2\sum_n\tau \mr{Im}\Bigl[\m{Q}^{\alpha\beta}_{D;n}\m{J}^{\gamma}_n\Bigl(-\frac{\p f}{\p\omega}\Bigr) + (\beta\leftrightarrow\gamma)\Bigr]_{\omega=\epsilon_n + i\eta}\label{BCD2}\\
        \tilde{\sigma}^{\alpha;\beta\gamma}_{\mr{Inter}}&=&2 \sum_{nm(m\neq n)}\mr{Re}\Bigl[ \frac{\m{J}^{\alpha}_{nm}\m{J}^{\beta\gamma}_{mn}}{(\epsilon_{nm}+2i\eta)^2}\Bigl(-\frac{\p f}{\p\omega}\Bigr) + \sum_{l\neq n} \frac{\m{J}^{\alpha}_{nm}\m{J}^{\beta}_{ml}\m{J}^{\gamma}_{ln} + (\beta\leftrightarrow\gamma)}{(\epsilon_{nm}\!+\!2i\eta)^2(\epsilon_{nl}\!+\!2i\eta)}\Bigl(-\frac{\p f}{\p\omega}\Bigr)\Bigr]_{\omega=\epsilon_n+i\eta}\label{Inter2}
    \end{eqnarray}
\end{widetext}
 $\sigma^{\alpha;\beta\gamma}_{\mr{M}}$ is the Matsubara term for the second order nonlinear response. Tilde above the terms means the classification of terms in the limit $\eta\rightarrow0$. First, we analyze the "Drude term" $\tilde{\sigma}^{\alpha;\beta\gamma}_{\mr{Drude}}$, which corresponds to the Drude term in the conventional analysis. While the first term and second term in Eq.~(\ref{Drude2}) only get the contribution from the real part of the distribution function, the third term has the contribution from the imaginary part. Under the approximation $\mr{Re}f(\epsilon_n+i\eta)\simeq f(\epsilon_n)$ and $\mr{Im}f(\epsilon_n+i\eta)\simeq \eta \p f(\epsilon_n)/\p\omega$, we can derive the following form from the third term in Eq.~(\ref{Drude2}) as,
\begin{eqnarray}
   &&\sum_{m(\neq n)}\m{J}^{\alpha}_n\frac{\m{J}^{\beta}_{nm}\m{J}^{\gamma}_{mn} + (\beta\leftrightarrow\gamma)}{\epsilon_{nm} + 2i\eta}\Bigl(-\frac{\p f}{\p\omega}\Bigr)\Bigr]_{\omega=\epsilon_n+i\eta}\nonumber\\
   &&=2\sum_{n\neq m}\tau^2\epsilon_{nm}\m{J}^{\alpha}_n g^{\beta\gamma}_{S;n,m}\mr{Re}\Bigl(\frac{\p f}{\p\omega}\Bigr)_{\omega=\epsilon_n+i\eta}\nonumber\\
   && \ \ \ \ + 2\sum_n\tau\m{J}^{\alpha}_n g^{\beta\gamma}_{S;n}\mr{Im}\Bigl(\frac{\p f}{\p\omega}\Bigr)_{\omega=\epsilon_n+i\eta}\\
   &&\simeq 2\sum_{n\neq m}\tau^2\epsilon_{nm}\m{J}^{\alpha}_n g^{\beta\gamma}_{S;n,m}\Bigl(\frac{\p f}{\p\omega}\Bigr)_{\omega=\epsilon_n}\nonumber\\
   && \ \ \ \ + \sum_n\m{J}^{\alpha}_n g^{\beta\gamma}_{S;n}\Bigl(\frac{\p^2 f}{\p\omega^2}\Bigr)_{\omega=\epsilon_n}
\end{eqnarray}
The second term newly arise from the imaginary part of the distribution function, which represents the broadening of the spectral function. We note that the other terms in $\tilde{\sigma}_{\mr{Drude}}$ can be summarized into the conventional Drude term as $\tau^2 \sum_n\m{J}^{\alpha}_n\p^{\beta}\p^{\gamma}f(\epsilon_n)$ in the limit of $\eta\rightarrow 0$. We can transform the second term into
\begin{eqnarray}
\tilde{\sigma}^{\alpha;\beta\gamma}_{\mr{Drude}}&=& \sigma^{\alpha;\beta\gamma}_{\mr{Drude}} + \sigma^{\alpha;\beta\gamma}_{\mr{sQMD}}\\
\sigma^{\alpha;\beta\gamma}_{\mr{Drude}} &=& 2\tau^2\sum_n\m{J}^{\alpha}_{n}\Biggl[\m{J}^{\beta}_{n}\m{J}^{\gamma}_{n}\mr{Re}\Bigl(-\frac{\p^2 f}{\p\omega^2}\Bigr) + \nonumber\\
&& \ \ \Bigl(\m{J}^{\beta\gamma}_{n}+ \sum_m\epsilon_{nm}g^{\beta\gamma}_{S;n,m}\Bigr)\Bigl(-\frac{\p f}{\p\omega}\Bigr)\Biggr]|_{\epsilon_n+i\eta}\nonumber\\
&\simeq& 2\tau^2 \sum_n\m{J}^{\alpha}_n\p^{\beta}\p^{\gamma}f(\epsilon_n)\\
\sigma^{\alpha;\beta\gamma}_{\mr{sQMD}} &=&2\sum_n\tau\m{J}^{\alpha}_n g^{\beta\gamma}_{S;n}\mr{Im}\Bigl(\frac{\p f}{\p\omega}\Bigr)_{\omega=\epsilon_n+i\eta}\nonumber\\
&\simeq&\sum_n\m{J}^{\alpha}_n g^{\beta\gamma}_{S;n}\Bigl(\frac{\p^2 f}{\p\omega^2}\Bigr)_{\omega=\epsilon_n}\nonumber\\
&=& \sum_n\p^{\alpha} g^{\beta\gamma}_{S;n}\Bigl(-\frac{\p f}{\p\omega}\Bigr)_{\omega=\epsilon_n},\label{sQMD}
\end{eqnarray}
and therefore, we call it the (smeared) quantum metric dipole term. Because here it is the smeared quantum metric, when there is a band-degeneracy at the Fermi surface, this term is proportional to $\tau^2$ (because $g^{\beta\gamma}_{S;n}$ is proportional to $\tau^2$ when $\epsilon_{nm}=0$) and gives the multi-band correction to the conventional nonlinear Drude term, as same as the linear conductivity. On the other hand, in the regime $|\epsilon_n-\epsilon_m|\tau\gg1$, this term becomes almost independent from the strength of the dissipation even through this term stems from the dissipation. 

Next, we analyze the originally Berry curvature dipole term. Under the approximation, we can derive 
\begin{eqnarray}
   \tilde{\sigma}^{\alpha;\beta\gamma}_{\mr{BCD}} &=&  \sigma^{\alpha;\beta\gamma}_{\mr{dBCD}} + \sigma^{\alpha;\beta\gamma}_{\mr{dQMD}}\\
   \sigma^{\alpha;\beta\gamma}_{\mr{dBCD}} &=& 2\sum_n \tau\Omega^{\alpha\beta}_{D;n}\m{J}^{\gamma}_n \mr{Re}\Bigl(-\frac{\p f}{\p\omega}\Bigr)_{\epsilon_n\!+\!i\eta} + (\beta \leftrightarrow \gamma)\nonumber\\
   &=&4\tau\sum_n \p^{\gamma}\Omega^{\alpha\beta}_{D;n}\mr{Re}f(\epsilon_n+i\eta) + (\beta \leftrightarrow \gamma)\label{dBCD}\\
   \sigma^{\alpha;\beta\gamma}_{\mr{dQMD}} &=& -2\sum_n g^{\alpha\beta}_{D;n}\m{J}^{\gamma}_n \mr{Im}\Bigl(-\frac{\p f}{\p\omega}\Bigr)_{\epsilon_n} + (\beta \leftrightarrow \gamma)\nonumber\\
   &=& 4\tau\sum_n \p^{\gamma}g^{\alpha\beta}_{D;n}\mr{Im}f(\epsilon_n+i\eta) + (\beta \leftrightarrow \gamma),\label{dQMD}
\end{eqnarray}
where $\Omega^{\alpha\beta}_{D;n} = \mr{Im}\m{Q}^{\alpha\beta}_{D;n}$ is the dissipative Berry curvature and $\Omega^{\alpha\beta}_{D;n} = 2\mr{Re}\m{Q}^{\alpha\beta}_{D;n}$ is the dissipative quantum metric.
 In the limit $\eta\rightarrow0$, Eq.~(\ref{dBCD}) becomes the well known the Berry curvature dipole term. We note that the dissipative quantum geometry is different from the smeared quantum geometry.  In this representation, because it is the dissipative geometry, the (dissipative) Berry curvature dipole term is not necessarily the Hall conductivity. Therefore, it is more convenient to convert the smeared geometric terms as written as,
 \begin{align}
    &\sigma^{\alpha;\beta\gamma}_{\mr{dBCD}}\nonumber\\
    &= 2\tau\sum_{n,m} \Biggl[\Bigl(\frac{\epsilon^2_{nm}\tau^2 -1}{\epsilon^2_{nm}\tau^2 +1} \Omega^{\alpha\beta}_{S;n,m} - \frac{2\epsilon_{nm}\tau}{\epsilon^2_{nm}\tau^2 +1}g^{\alpha\beta}_{S;n,m}\Bigr)\nonumber\\
    & \ \ \ \ \ \ \ \ \ \times\m{J}^{\gamma}_n\mr{Re}\Bigl(-\frac{\p f}{\p\omega}\Bigr)_{\epsilon_n\!+\!i\eta}+ (\beta \leftrightarrow \gamma)\Biggr]\label{dBCD_sQG}\\
    &\simeq 2\tau\sum_{n,m}\Biggl[\Bigl(\Omega^{\alpha\beta}_{S;n} - \frac{2g^{\alpha\beta}_{S;n,m}}{\epsilon_{nm}\tau}\Bigr)\m{J}^{\gamma}_n\Bigl(-\frac{\p f}{\p\omega}\Bigr)_{\epsilon_n} + (\beta \leftrightarrow \gamma)\Biggr]
\end{align}
\begin{align}
    &\sigma^{\alpha;\beta\gamma}_{\mr{dQMD}}\nonumber\\
    &= 2\tau\sum_{n,m}\Biggl[\Bigl(\frac{\epsilon^2_{nm}\tau^2 -1}{\epsilon^2_{nm}\tau^2+1} g^{\alpha\beta}_{S;n,m} + \frac{2\epsilon_{nm}\tau}{\epsilon^2_{nm}\tau^2 +1}\Omega^{\alpha\beta}_{S;n,m}\Bigr)\nonumber\\
    & \ \ \ \ \ \ \ \ \ \times\m{J}^{\gamma}_n\mr{Im}\Bigl(-\frac{\p f}{\p\omega}\Bigr)_{\epsilon_n\!+\!i\eta}+ (\beta \leftrightarrow \gamma)\Biggr]\label{dQMD_sQG}\\
    &\simeq \sum_{n}\Biggl[-\Bigl(\p^{\gamma}g^{\alpha\beta}_{S;n}\Bigr)\Bigl(-\frac{\p f}{\p\omega}\Bigr)_{\epsilon_n} + \frac{2\Omega^{\alpha\beta}_{S;n,m}}{\epsilon_{nm}\tau}\m{J}^{\gamma}_n\Bigl(-\frac{\p^2 f}{\p\omega^2}\Bigr)_{\epsilon_n}\nonumber\\
    & \ \ \ \ \ \ \ \ \ + (\beta \leftrightarrow \gamma)\Biggr]\label{dQMD_app}
 \end{align}
 In the limit $\epsilon_{nm}\tau\gg1$, up to the first order of $\eta$, the first term in Eq.~(\ref{dBCD_sQG}) is the conventional Berry curvature dipole term and the second term is the terms pointed out in Ref.\cite{PhysRevLett.112.166601} (the second term in Eq.~(13) in Ref.\cite{PhysRevLett.112.166601})
Here, another QMD term emerges from the original Berry curvature dipole term. Because it is the dissipative quantum metric, when there is a band-degeneracy at the Fermi surface, this term also gives the multi-band correction to the nonlinear Drude term. As same as the smeared QMD term, in the limit $|\epsilon_n-\epsilon_m|\tau\gg1$, this behaves as is almost independent of the strength of the dissipation, while it stems from the dissipation. 
When considering the limit $\epsilon_{nm}\tau \gg 1$ and approximating $\mr{Im}f(\epsilon_n+i\eta) = \eta(\p f/\p\omega)|_{\epsilon_n}$, which is enough justified at $\beta\eta<1$, the sum of the smeared QMD term and the first term in Eq.~(\ref{dQMD_app}) can be written as,
\begin{eqnarray}
   \Bigl(\p^{\gamma}g^{\alpha\beta}_{S;n}\!+\!\p^{\beta}g^{\alpha\gamma}_{S;n}\!-\!\p^{\alpha}g^{\beta\gamma}_{S;n}\Bigr)\Bigl(\frac{\p f}{\p\omega}\Bigr)_{\epsilon_n}= 2 \Gamma^{\alpha;\beta\gamma}_{S;n} \Bigl(\frac{\p f}{\p\omega}\Bigr)_{\epsilon_n},\nonumber\\
\end{eqnarray}
where $\Gamma^{\alpha;\beta\gamma}_{S;n} \equiv (\p^{\gamma}g^{\alpha\beta}_{S;n} + \p^{\beta}g^{\alpha\gamma}_{S;n} - \p^{\alpha}g^{\beta\gamma}_{S;n})/2$ can be called the ``smeared Christoffel symbol'', which is the generalization of the Christoffel symbol in Ref.\cite{PhysRevX.10.041041} to the dissipative case. Therefore, we can call the sum of those term as ``the smeared Christoffel symbol term'' at the Fermi surface.

Finally, we analyze the interband term $\tilde{\sigma}^{\alpha;\beta\gamma}_{\mr{Inter}}$ in Eq.~(\ref{Inter2}).

$\tilde{\sigma}^{\alpha;\beta\gamma}_{\mr{Inter}}$ can generate the another Christoffel symbol term and the generalized Berry curvature term  as,
\begin{eqnarray}
   &&\sigma^{\alpha;\beta\gamma}_{\mr{Inter;1}} = \sigma^{\alpha;\beta\gamma}_{\mr{ChS:I\hspace{-.01em}I}} + \sigma^{\alpha;\beta\gamma}_{\mr{gBC}},\\
   &&\sigma^{\alpha;\beta\gamma}_{\mr{ChS:I\hspace{-.01em}I}}\nonumber\\ 
   &&= 2\sum_{\bk}\sum_{n,m(\neq n)}\Bigl[ \mr{Re}(\m{J}^{\alpha}_{nm}\m{J}^{\beta\gamma}_{mn})\mr{Re}\Bigl(\frac{1}{(\epsilon_{nm}\!+\!2i\eta)^2}\Bigr)\nonumber\\
   && \ \  + \Bigl\{\sum_{l(\neq n)}\mr{Re}(\m{J}^{\alpha}_{nm}\m{J}^{\beta}_{ml}\m{J}^{\gamma}_{ln})\mr{Re}\Bigl(\frac{1}{(\epsilon_{nm}\!+\!2i\eta)^2(\epsilon_{nl}\!+\!2i\eta)}\Bigr)\Bigr\}\nonumber\\
   && \ \  + \Bigl\{\beta\leftrightarrow\gamma\Bigr\} \Bigr] \mr{Re}\Bigl(-\frac{\p f}{\p\omega}\Bigr)_{\epsilon_n+i\eta} + \m{O}(\tau^{-2})\label{app:ChS2}\\
   &&= 2\sum_{\bk}\sum_{n}g^{\alpha;\beta\gamma}_{S';n}\Bigl(-\frac{\p f}{\p\omega}\Bigr)_{\epsilon_n} = 2\sum_{\bk}\sum_{n}\Gamma^{\alpha;\beta\gamma}_{S';n}\Bigl(-\frac{\p f}{\p\omega}\Bigr)_{\epsilon_n}\nonumber\\
   &&\\
   &&\simeq \sum_{\bk}\sum_n \Bigl(\braket{\p^{\alpha}n|\p^{\beta\gamma}n} +\braket{\p^{\beta\gamma}n|\p^{\alpha}n} \Bigr)\Bigl(-\frac{\p f}{\p\omega}\Bigr)_{\epsilon_n}\label{app:ChS2_app}\\
   &&= 2\sum_{\bk}\sum_{n}\Gamma^{\alpha;\beta\gamma}_{n}\Bigl(-\frac{\p f}{\p\omega}\Bigr)_{\epsilon_n}\\
   &&\Gamma^{\alpha;\beta\gamma}_{n} = \frac{1}{2}\Bigl(\p^{\gamma}g^{\alpha\beta}_{n}+\p^{\beta}g^{\gamma\alpha}_{n}-\p^{\alpha}g^{\beta\gamma}_{n}\Bigr)\nonumber\\
   && \ \ \ \ \ \ \ \ =\frac{1}{2}\Bigl(\braket{\p^{\alpha}n|\p^{\beta\gamma}n} +\braket{\p^{\beta\gamma}n|\p^{\alpha}n} \Bigr)
  \end{eqnarray}
\begin{eqnarray}
   &&\sigma^{\alpha;\beta\gamma}_{\mr{gBC}} = \sigma^{\alpha;\beta\gamma}_{\mr{gBC:re}} + \sigma^{\alpha;\beta\gamma}_{\mr{gBC:im}} + \sigma^{\alpha;\beta\gamma}_{\mr{gBC:add}}\label{app:gBC}\\
   &&\sigma^{\alpha;\beta\gamma}_{\mr{gBC:re}}+ \sigma^{\alpha;\beta\gamma}_{\mr{gBC:add}}\nonumber\\
   &&= 2\sum_{\bk}\sum_{n,m(\neq n)}\Bigl[ \mr{Im}(\m{J}^{\alpha}_{nm}\m{J}^{\beta\gamma}_{mn})\mr{Im}\Bigl(\frac{1}{(\epsilon_{nm}\!+\!2i\eta)^2}\Bigr)\nonumber\\
   && \ \  + \Bigl\{\sum_{l(\neq n)}\mr{Im}(\m{J}^{\alpha}_{nm}\m{J}^{\beta}_{ml}\m{J}^{\gamma}_{ln})\mr{Im}\Bigl(\frac{1}{(\epsilon_{nm}\!+\!2i\eta)^2(\epsilon_{nl}\!+\!2i\eta)}\Bigr)\Bigr\}\nonumber\\
   && \ \  + \Bigl\{\beta\leftrightarrow\gamma\Bigr\} \Bigr] \mr{Re}\Bigl(\frac{\p f}{\p\omega}\Bigr)_{\epsilon_n+i\eta}\label{app:gBCD}\\
   &&= \sum_{\bk}\sum_{n,m(\neq n)}\frac{\Omega^{\alpha,\beta\gamma}_{S';n,m}}{\epsilon_{nm}\tau}\Bigl(-\frac{\p f}{\p\omega}\Bigr)_{\epsilon_n}+ \sigma^{\alpha;\beta\gamma}_{\mr{gBC:add}} + \m{O}(\tau^{-2}),\label{app:gBCre}\\
   &&\sigma^{\alpha;\beta\gamma}_{\mr{gBC:add}}\nonumber\\
   &&\simeq2\sum_{\bk}\sum_{n,m,l(\neq n)}\frac{\mr{Im}(\m{J}^{\alpha}_{nm}\m{J}^{\beta}_{ml}\m{J}^{\gamma}_{ln})}{\epsilon^2_{nm}\epsilon_{nl}}\Bigl(\frac{1}{\epsilon_{nm}\tau}-\frac{1}{\epsilon_{nl}\tau}\Bigr)\Bigl(-\frac{\p f}{\p\omega}\Bigr)_{\epsilon_n},\nonumber\\
   && \ \ \ \ \ \ \ \ \ \ \ \ \ \ \ \ \ \ \ + (\beta\leftrightarrow\gamma)\label{app:gBCadd}
\end{eqnarray}
\begin{eqnarray}
   &&\sigma^{\alpha;\beta\gamma}_{\mr{gBC:im}}\nonumber\\
   &&= 2\sum_{\bk}\sum_{n,m(\neq n)}\Bigl[ \mr{Im}(\m{J}^{\alpha}_{nm}\m{J}^{\beta\gamma}_{mn})\mr{Re}\Bigl(\frac{1}{(\epsilon_{nm}\!+\!2i\eta)^2}\Bigr)\nonumber\\
   && \ \  + \Bigl\{\sum_{l(\neq n)}\mr{Im}(\m{J}^{\alpha}_{nm}\m{J}^{\beta}_{ml}\m{J}^{\gamma}_{ln})\mr{Re}\Bigl(\frac{1}{(\epsilon_{nm}\!+\!2i\eta)^2(\epsilon_{nl}\!+\!2i\eta)}\Bigr)\Bigr\}\nonumber\\
   && \ \  + \Bigl\{\beta\leftrightarrow\gamma\Bigr\} \Bigr] \mr{Im}\Bigl(\frac{\p f}{\p\omega}\Bigr)_{\epsilon_n+i\eta}\label{app:gBCDim}\\
   &&= -\sum_{\bk}\sum_{n}\frac{\Omega^{\alpha,\beta\gamma}_{S';n}}{\tau}\Bigl(-\frac{\p^2 f}{\p\omega^2}\Bigr)_{\epsilon_n} + \m{O}(\tau^2),\label{app:gBCim}
\end{eqnarray}
where the $\Omega^{\alpha,\beta\gamma}_{S';n,m} (g^{\alpha,\beta\gamma}_{S';n,m})$ is the smeared Berry curvature (quantum metric) generalized to the higher-order derivative, which can be derived by the following substitution from $\Omega^{\alpha,\beta\gamma}_{n,m} (g^{\alpha,\beta\gamma}_{n,m})$,
\begin{eqnarray}
    &&\begin{dcases}
   \frac{1}{\epsilon^2_{nm}}\rightarrow \frac{1}{\epsilon^2_{nm}+4\eta^2}\\
   \frac{1}{\epsilon_{nm}}\rightarrow \frac{\epsilon_{nm}}{\epsilon^2_{nm}+4\eta^2} ,
   \end{dcases}\\
    g^{\alpha,\beta\gamma}_{n,m} &=&\mr{Re}\braket{\p^\alpha n|m}\braket{m|\p^{\beta\gamma}n} = \Gamma^{\alpha;\beta\gamma}_{n,m}\\ &=&2\mr{Re}\Bigl[\frac{\m{J}^{\alpha}_{nm}}{\epsilon_{nm}}\frac{1}{\epsilon_{nm}}\Bigl(\m{J}^{\beta\gamma}_{mn} + \frac{\m{J}^{\beta}_{ml}\m{J}^{\gamma}_{ln}+(\beta\leftrightarrow\gamma)}{\epsilon_{nl}}\Bigr)\Bigr]\nonumber\\
    &&\\
   \Omega^{\alpha,\beta\gamma}_{n,m} &=& 2\mr{Im}\braket{\p^\alpha n|m}\braket{m|\p^{\beta\gamma}n}
\end{eqnarray}
If we consider $\p^{\beta\gamma}$ as the derivative to the new direction, $\Omega^{\alpha,\beta\gamma}_{S';n,m} (g^{\alpha,\beta\gamma}_{S';n,m})$ is also the smeared Berry curvature (quantum metric) generalized to the higher-order derivative.
We note that, in two-band systems under the time-reversal symmetry, only the first term in Eq.~(\ref{app:gBCD}) can induce the non-reciprocal conductivity $\sigma^{\alpha;\alpha\alpha}$. Under the time-reversal symmetry, the Drude term, the smeared Christoffel term and the other terms proportional to the smeared quantum metric are zero. Moreover, the terms proportional to the conventional smeared Berry curvature is also zero because $\alpha=\beta=\gamma$. However, the extended Berry curvature $\Omega^{\alpha,\alpha\alpha}$ can be finite because $\p^{\alpha\alpha}$ is different from $\p^{\alpha}$. 
As we can see in Eq.~(\ref{app:gBCre}), Eq.~(\ref{app:gBCadd}) and Eq.~(\ref{app:gBCim}), this non-reciprocal term under the time-reversal symmetry becomes zero when $\eta\rightarrow0$, which is pointed out in Ref.\cite{Morimoto2018}. Here we also find another necessary condition, which is that the Hamiltonian must have the quadratic term, because $\m{J}^{\alpha\alpha}$ becomes zero without it. Therefore, non-reciprocal transport under the time-reversal symmetry in DC limit is zero, for example in the Weyl system which has the linear dispersion.

\bibliography{main.bib}

\begin{thebibliography}{50}%
\makeatletter
\providecommand \@ifxundefined [1]{%
 \@ifx{#1\undefined}
}%
\providecommand \@ifnum [1]{%
 \ifnum #1\expandafter \@firstoftwo
 \else \expandafter \@secondoftwo
 \fi
}%
\providecommand \@ifx [1]{%
 \ifx #1\expandafter \@firstoftwo
 \else \expandafter \@secondoftwo
 \fi
}%
\providecommand \natexlab [1]{#1}%
\providecommand \enquote  [1]{``#1''}%
\providecommand \bibnamefont  [1]{#1}%
\providecommand \bibfnamefont [1]{#1}%
\providecommand \citenamefont [1]{#1}%
\providecommand \href@noop [0]{\@secondoftwo}%
\providecommand \href [0]{\begingroup \@sanitize@url \@href}%
\providecommand \@href[1]{\@@startlink{#1}\@@href}%
\providecommand \@@href[1]{\endgroup#1\@@endlink}%
\providecommand \@sanitize@url [0]{\catcode `\\12\catcode `\$12\catcode
  `\&12\catcode `\#12\catcode `\^12\catcode `\_12\catcode `\%12\relax}%
\providecommand \@@startlink[1]{}%
\providecommand \@@endlink[0]{}%
\providecommand \url  [0]{\begingroup\@sanitize@url \@url }%
\providecommand \@url [1]{\endgroup\@href {#1}{\urlprefix }}%
\providecommand \urlprefix  [0]{URL }%
\providecommand \Eprint [0]{\href }%
\providecommand \doibase [0]{http://dx.doi.org/}%
\providecommand \selectlanguage [0]{\@gobble}%
\providecommand \bibinfo  [0]{\@secondoftwo}%
\providecommand \bibfield  [0]{\@secondoftwo}%
\providecommand \translation [1]{[#1]}%
\providecommand \BibitemOpen [0]{}%
\providecommand \bibitemStop [0]{}%
\providecommand \bibitemNoStop [0]{.\EOS\space}%
\providecommand \EOS [0]{\spacefactor3000\relax}%
\providecommand \BibitemShut  [1]{\csname bibitem#1\endcsname}%
\let\auto@bib@innerbib\@empty
\bibitem [{\citenamefont {Petersen}\ \emph {et~al.}(2006)\citenamefont
  {Petersen}, \citenamefont {Caswell}, \citenamefont {Dodge}, \citenamefont
  {Sergienko}, \citenamefont {He}, \citenamefont {Jin},\ and\ \citenamefont
  {Mandrus}}]{Petersen2006}%
  \BibitemOpen
  \bibfield  {author} {\bibinfo {author} {\bibfnamefont {J.~C.}\ \bibnamefont
  {Petersen}}, \bibinfo {author} {\bibfnamefont {M.~D.}\ \bibnamefont
  {Caswell}}, \bibinfo {author} {\bibfnamefont {J.~S.}\ \bibnamefont {Dodge}},
  \bibinfo {author} {\bibfnamefont {I.~A.}\ \bibnamefont {Sergienko}}, \bibinfo
  {author} {\bibfnamefont {J.}~\bibnamefont {He}}, \bibinfo {author}
  {\bibfnamefont {R.}~\bibnamefont {Jin}}, \ and\ \bibinfo {author}
  {\bibfnamefont {D.}~\bibnamefont {Mandrus}},\ }\href {\doibase
  10.1038/nphys392} {\bibfield  {journal} {\bibinfo  {journal} {Nature
  Physics}\ }\textbf {\bibinfo {volume} {2}},\ \bibinfo {pages} {605} (\bibinfo
  {year} {2006})}\BibitemShut {NoStop}%
\bibitem [{\citenamefont {Zhao}\ \emph {et~al.}(2016)\citenamefont {Zhao},
  \citenamefont {Torchinsky}, \citenamefont {Chu}, \citenamefont {Ivanov},
  \citenamefont {Lifshitz}, \citenamefont {Flint}, \citenamefont {Qi},
  \citenamefont {Cao},\ and\ \citenamefont {Hsieh}}]{Zhao2016}%
  \BibitemOpen
  \bibfield  {author} {\bibinfo {author} {\bibfnamefont {L.}~\bibnamefont
  {Zhao}}, \bibinfo {author} {\bibfnamefont {D.~H.}\ \bibnamefont
  {Torchinsky}}, \bibinfo {author} {\bibfnamefont {H.}~\bibnamefont {Chu}},
  \bibinfo {author} {\bibfnamefont {V.}~\bibnamefont {Ivanov}}, \bibinfo
  {author} {\bibfnamefont {R.}~\bibnamefont {Lifshitz}}, \bibinfo {author}
  {\bibfnamefont {R.}~\bibnamefont {Flint}}, \bibinfo {author} {\bibfnamefont
  {T.}~\bibnamefont {Qi}}, \bibinfo {author} {\bibfnamefont {G.}~\bibnamefont
  {Cao}}, \ and\ \bibinfo {author} {\bibfnamefont {D.}~\bibnamefont {Hsieh}},\
  }\href {\doibase 10.1038/nphys3517} {\bibfield  {journal} {\bibinfo
  {journal} {Nature Physics}\ }\textbf {\bibinfo {volume} {12}},\ \bibinfo
  {pages} {32} (\bibinfo {year} {2016})}\BibitemShut {NoStop}%
\bibitem [{\citenamefont {Harter}\ \emph {et~al.}(2017)\citenamefont {Harter},
  \citenamefont {Zhao}, \citenamefont {Yan}, \citenamefont {Mandrus},\ and\
  \citenamefont {Hsieh}}]{Harter295}%
  \BibitemOpen
  \bibfield  {author} {\bibinfo {author} {\bibfnamefont {J.~W.}\ \bibnamefont
  {Harter}}, \bibinfo {author} {\bibfnamefont {Z.~Y.}\ \bibnamefont {Zhao}},
  \bibinfo {author} {\bibfnamefont {J.-Q.}\ \bibnamefont {Yan}}, \bibinfo
  {author} {\bibfnamefont {D.~G.}\ \bibnamefont {Mandrus}}, \ and\ \bibinfo
  {author} {\bibfnamefont {D.}~\bibnamefont {Hsieh}},\ }\href {\doibase
  10.1126/science.aad1188} {\bibfield  {journal} {\bibinfo  {journal}
  {Science}\ }\textbf {\bibinfo {volume} {356}},\ \bibinfo {pages} {295}
  (\bibinfo {year} {2017})}\BibitemShut {NoStop}%
\bibitem [{\citenamefont {Sipe}\ and\ \citenamefont
  {Ghahramani}(1993)}]{PhysRevB.48.11705}%
  \BibitemOpen
  \bibfield  {author} {\bibinfo {author} {\bibfnamefont {J.~E.}\ \bibnamefont
  {Sipe}}\ and\ \bibinfo {author} {\bibfnamefont {E.}~\bibnamefont
  {Ghahramani}},\ }\href {\doibase 10.1103/PhysRevB.48.11705} {\bibfield
  {journal} {\bibinfo  {journal} {Phys. Rev. B}\ }\textbf {\bibinfo {volume}
  {48}},\ \bibinfo {pages} {11705} (\bibinfo {year} {1993})}\BibitemShut
  {NoStop}%
\bibitem [{\citenamefont {Morimoto}\ and\ \citenamefont
  {Nagaosa}(2016)}]{Morimotoe1501524}%
  \BibitemOpen
  \bibfield  {author} {\bibinfo {author} {\bibfnamefont {T.}~\bibnamefont
  {Morimoto}}\ and\ \bibinfo {author} {\bibfnamefont {N.}~\bibnamefont
  {Nagaosa}},\ }\href {\doibase 10.1126/sciadv.1501524} {\bibfield  {journal}
  {\bibinfo  {journal} {Science Advances}\ }\textbf {\bibinfo {volume} {2}}
  (\bibinfo {year} {2016}),\ 10.1126/sciadv.1501524}\BibitemShut {NoStop}%
\bibitem [{\citenamefont {Wu}\ \emph {et~al.}(2017)\citenamefont {Wu},
  \citenamefont {Patankar}, \citenamefont {Morimoto}, \citenamefont {Nair},
  \citenamefont {Thewalt}, \citenamefont {Little}, \citenamefont {Analytis},
  \citenamefont {Moore},\ and\ \citenamefont {Orenstein}}]{Wu2017}%
  \BibitemOpen
  \bibfield  {author} {\bibinfo {author} {\bibfnamefont {L.}~\bibnamefont
  {Wu}}, \bibinfo {author} {\bibfnamefont {S.}~\bibnamefont {Patankar}},
  \bibinfo {author} {\bibfnamefont {T.}~\bibnamefont {Morimoto}}, \bibinfo
  {author} {\bibfnamefont {N.~L.}\ \bibnamefont {Nair}}, \bibinfo {author}
  {\bibfnamefont {E.}~\bibnamefont {Thewalt}}, \bibinfo {author} {\bibfnamefont
  {A.}~\bibnamefont {Little}}, \bibinfo {author} {\bibfnamefont {J.~G.}\
  \bibnamefont {Analytis}}, \bibinfo {author} {\bibfnamefont {J.~E.}\
  \bibnamefont {Moore}}, \ and\ \bibinfo {author} {\bibfnamefont
  {J.}~\bibnamefont {Orenstein}},\ }\href {\doibase 10.1038/nphys3969}
  {\bibfield  {journal} {\bibinfo  {journal} {Nature Physics}\ }\textbf
  {\bibinfo {volume} {13}},\ \bibinfo {pages} {350} (\bibinfo {year}
  {2017})}\BibitemShut {NoStop}%
\bibitem [{\citenamefont {Sodemann}\ and\ \citenamefont
  {Fu}(2015)}]{PhysRevLett.115.216806}%
  \BibitemOpen
  \bibfield  {author} {\bibinfo {author} {\bibfnamefont {I.}~\bibnamefont
  {Sodemann}}\ and\ \bibinfo {author} {\bibfnamefont {L.}~\bibnamefont {Fu}},\
  }\href {\doibase 10.1103/PhysRevLett.115.216806} {\bibfield  {journal}
  {\bibinfo  {journal} {Phys. Rev. Lett.}\ }\textbf {\bibinfo {volume} {115}},\
  \bibinfo {pages} {216806} (\bibinfo {year} {2015})}\BibitemShut {NoStop}%
\bibitem [{\citenamefont {Ma}\ \emph {et~al.}(2019)\citenamefont {Ma},
  \citenamefont {Xu}, \citenamefont {Shen}, \citenamefont {MacNeill},
  \citenamefont {Fatemi}, \citenamefont {Chang}, \citenamefont {Mier~Valdivia},
  \citenamefont {Wu}, \citenamefont {Du}, \citenamefont {Hsu}, \citenamefont
  {Fang}, \citenamefont {Gibson}, \citenamefont {Watanabe}, \citenamefont
  {Taniguchi}, \citenamefont {Cava}, \citenamefont {Kaxiras}, \citenamefont
  {Lu}, \citenamefont {Lin}, \citenamefont {Fu}, \citenamefont {Gedik},\ and\
  \citenamefont {Jarillo-Herrero}}]{Ma2019}%
  \BibitemOpen
  \bibfield  {author} {\bibinfo {author} {\bibfnamefont {Q.}~\bibnamefont
  {Ma}}, \bibinfo {author} {\bibfnamefont {S.-Y.}\ \bibnamefont {Xu}}, \bibinfo
  {author} {\bibfnamefont {H.}~\bibnamefont {Shen}}, \bibinfo {author}
  {\bibfnamefont {D.}~\bibnamefont {MacNeill}}, \bibinfo {author}
  {\bibfnamefont {V.}~\bibnamefont {Fatemi}}, \bibinfo {author} {\bibfnamefont
  {T.-R.}\ \bibnamefont {Chang}}, \bibinfo {author} {\bibfnamefont {A.~M.}\
  \bibnamefont {Mier~Valdivia}}, \bibinfo {author} {\bibfnamefont
  {S.}~\bibnamefont {Wu}}, \bibinfo {author} {\bibfnamefont {Z.}~\bibnamefont
  {Du}}, \bibinfo {author} {\bibfnamefont {C.-H.}\ \bibnamefont {Hsu}},
  \bibinfo {author} {\bibfnamefont {S.}~\bibnamefont {Fang}}, \bibinfo {author}
  {\bibfnamefont {Q.~D.}\ \bibnamefont {Gibson}}, \bibinfo {author}
  {\bibfnamefont {K.}~\bibnamefont {Watanabe}}, \bibinfo {author}
  {\bibfnamefont {T.}~\bibnamefont {Taniguchi}}, \bibinfo {author}
  {\bibfnamefont {R.~J.}\ \bibnamefont {Cava}}, \bibinfo {author}
  {\bibfnamefont {E.}~\bibnamefont {Kaxiras}}, \bibinfo {author} {\bibfnamefont
  {H.-Z.}\ \bibnamefont {Lu}}, \bibinfo {author} {\bibfnamefont
  {H.}~\bibnamefont {Lin}}, \bibinfo {author} {\bibfnamefont {L.}~\bibnamefont
  {Fu}}, \bibinfo {author} {\bibfnamefont {N.}~\bibnamefont {Gedik}}, \ and\
  \bibinfo {author} {\bibfnamefont {P.}~\bibnamefont {Jarillo-Herrero}},\
  }\href {\doibase 10.1038/s41586-018-0807-6} {\bibfield  {journal} {\bibinfo
  {journal} {Nature}\ }\textbf {\bibinfo {volume} {565}},\ \bibinfo {pages}
  {337} (\bibinfo {year} {2019})}\BibitemShut {NoStop}%
\bibitem [{\citenamefont {Kang}\ \emph {et~al.}(2019)\citenamefont {Kang},
  \citenamefont {Li}, \citenamefont {Sohn}, \citenamefont {Shan},\ and\
  \citenamefont {Mak}}]{Kang2019}%
  \BibitemOpen
  \bibfield  {author} {\bibinfo {author} {\bibfnamefont {K.}~\bibnamefont
  {Kang}}, \bibinfo {author} {\bibfnamefont {T.}~\bibnamefont {Li}}, \bibinfo
  {author} {\bibfnamefont {E.}~\bibnamefont {Sohn}}, \bibinfo {author}
  {\bibfnamefont {J.}~\bibnamefont {Shan}}, \ and\ \bibinfo {author}
  {\bibfnamefont {K.~F.}\ \bibnamefont {Mak}},\ }\href {\doibase
  10.1038/s41563-019-0294-7} {\bibfield  {journal} {\bibinfo  {journal} {Nature
  Materials}\ }\textbf {\bibinfo {volume} {18}},\ \bibinfo {pages} {324}
  (\bibinfo {year} {2019})}\BibitemShut {NoStop}%
\bibitem [{\citenamefont {Dzsaber}\ \emph {et~al.}(2021)\citenamefont
  {Dzsaber}, \citenamefont {Yan}, \citenamefont {Taupin}, \citenamefont
  {Eguchi}, \citenamefont {Prokofiev}, \citenamefont {Shiroka}, \citenamefont
  {Blaha}, \citenamefont {Rubel}, \citenamefont {Grefe}, \citenamefont {Lai},
  \citenamefont {Si},\ and\ \citenamefont {Paschen}}]{Dzsabere2013386118}%
  \BibitemOpen
  \bibfield  {author} {\bibinfo {author} {\bibfnamefont {S.}~\bibnamefont
  {Dzsaber}}, \bibinfo {author} {\bibfnamefont {X.}~\bibnamefont {Yan}},
  \bibinfo {author} {\bibfnamefont {M.}~\bibnamefont {Taupin}}, \bibinfo
  {author} {\bibfnamefont {G.}~\bibnamefont {Eguchi}}, \bibinfo {author}
  {\bibfnamefont {A.}~\bibnamefont {Prokofiev}}, \bibinfo {author}
  {\bibfnamefont {T.}~\bibnamefont {Shiroka}}, \bibinfo {author} {\bibfnamefont
  {P.}~\bibnamefont {Blaha}}, \bibinfo {author} {\bibfnamefont
  {O.}~\bibnamefont {Rubel}}, \bibinfo {author} {\bibfnamefont {S.~E.}\
  \bibnamefont {Grefe}}, \bibinfo {author} {\bibfnamefont {H.-H.}\ \bibnamefont
  {Lai}}, \bibinfo {author} {\bibfnamefont {Q.}~\bibnamefont {Si}}, \ and\
  \bibinfo {author} {\bibfnamefont {S.}~\bibnamefont {Paschen}},\ }\href
  {\doibase 10.1073/pnas.2013386118} {\bibfield  {journal} {\bibinfo  {journal}
  {Proceedings of the National Academy of Sciences}\ }\textbf {\bibinfo
  {volume} {118}} (\bibinfo {year} {2021}),\ 10.1073/pnas.2013386118},\ \Eprint
  {http://arxiv.org/abs/https://www.pnas.org/content/118/8/e2013386118.full.pdf}
  {https://www.pnas.org/content/118/8/e2013386118.full.pdf} \BibitemShut
  {NoStop}%
\bibitem [{\citenamefont {Zhou}\ \emph {et~al.}(2020)\citenamefont {Zhou},
  \citenamefont {Zhang},\ and\ \citenamefont {Law}}]{PhysRevApplied.13.024053}%
  \BibitemOpen
  \bibfield  {author} {\bibinfo {author} {\bibfnamefont {B.~T.}\ \bibnamefont
  {Zhou}}, \bibinfo {author} {\bibfnamefont {C.-P.}\ \bibnamefont {Zhang}}, \
  and\ \bibinfo {author} {\bibfnamefont {K.}~\bibnamefont {Law}},\ }\href
  {\doibase 10.1103/PhysRevApplied.13.024053} {\bibfield  {journal} {\bibinfo
  {journal} {Phys. Rev. Applied}\ }\textbf {\bibinfo {volume} {13}},\ \bibinfo
  {pages} {024053} (\bibinfo {year} {2020})}\BibitemShut {NoStop}%
\bibitem [{\citenamefont {McIver}\ \emph {et~al.}(2012)\citenamefont {McIver},
  \citenamefont {Hsieh}, \citenamefont {Steinberg}, \citenamefont
  {Jarillo-Herrero},\ and\ \citenamefont {Gedik}}]{McIver2012}%
  \BibitemOpen
  \bibfield  {author} {\bibinfo {author} {\bibfnamefont {J.~W.}\ \bibnamefont
  {McIver}}, \bibinfo {author} {\bibfnamefont {D.}~\bibnamefont {Hsieh}},
  \bibinfo {author} {\bibfnamefont {H.}~\bibnamefont {Steinberg}}, \bibinfo
  {author} {\bibfnamefont {P.}~\bibnamefont {Jarillo-Herrero}}, \ and\ \bibinfo
  {author} {\bibfnamefont {N.}~\bibnamefont {Gedik}},\ }\href {\doibase
  10.1038/nnano.2011.214} {\bibfield  {journal} {\bibinfo  {journal} {Nature
  Nanotechnology}\ }\textbf {\bibinfo {volume} {7}},\ \bibinfo {pages} {96}
  (\bibinfo {year} {2012})}\BibitemShut {NoStop}%
\bibitem [{\citenamefont {Kastl}\ \emph {et~al.}(2015)\citenamefont {Kastl},
  \citenamefont {Karnetzky}, \citenamefont {Karl},\ and\ \citenamefont
  {Holleitner}}]{Kastl2015}%
  \BibitemOpen
  \bibfield  {author} {\bibinfo {author} {\bibfnamefont {C.}~\bibnamefont
  {Kastl}}, \bibinfo {author} {\bibfnamefont {C.}~\bibnamefont {Karnetzky}},
  \bibinfo {author} {\bibfnamefont {H.}~\bibnamefont {Karl}}, \ and\ \bibinfo
  {author} {\bibfnamefont {A.~W.}\ \bibnamefont {Holleitner}},\ }\href
  {\doibase 10.1038/ncomms7617} {\bibfield  {journal} {\bibinfo  {journal}
  {Nature Communications}\ }\textbf {\bibinfo {volume} {6}},\ \bibinfo {pages}
  {6617} (\bibinfo {year} {2015})}\BibitemShut {NoStop}%
\bibitem [{\citenamefont {Chan}\ \emph {et~al.}(2017)\citenamefont {Chan},
  \citenamefont {Lindner}, \citenamefont {Refael},\ and\ \citenamefont
  {Lee}}]{PhysRevB.95.041104}%
  \BibitemOpen
  \bibfield  {author} {\bibinfo {author} {\bibfnamefont {C.-K.}\ \bibnamefont
  {Chan}}, \bibinfo {author} {\bibfnamefont {N.~H.}\ \bibnamefont {Lindner}},
  \bibinfo {author} {\bibfnamefont {G.}~\bibnamefont {Refael}}, \ and\ \bibinfo
  {author} {\bibfnamefont {P.~A.}\ \bibnamefont {Lee}},\ }\href {\doibase
  10.1103/PhysRevB.95.041104} {\bibfield  {journal} {\bibinfo  {journal} {Phys.
  Rev. B}\ }\textbf {\bibinfo {volume} {95}},\ \bibinfo {pages} {041104}
  (\bibinfo {year} {2017})}\BibitemShut {NoStop}%
\bibitem [{\citenamefont {Cook}\ \emph {et~al.}(2017)\citenamefont {Cook},
  \citenamefont {M.~Fregoso}, \citenamefont {de~Juan}, \citenamefont {Coh},\
  and\ \citenamefont {Moore}}]{Cook2017}%
  \BibitemOpen
  \bibfield  {author} {\bibinfo {author} {\bibfnamefont {A.~M.}\ \bibnamefont
  {Cook}}, \bibinfo {author} {\bibfnamefont {B.}~\bibnamefont {M.~Fregoso}},
  \bibinfo {author} {\bibfnamefont {F.}~\bibnamefont {de~Juan}}, \bibinfo
  {author} {\bibfnamefont {S.}~\bibnamefont {Coh}}, \ and\ \bibinfo {author}
  {\bibfnamefont {J.~E.}\ \bibnamefont {Moore}},\ }\href {\doibase
  10.1038/ncomms14176} {\bibfield  {journal} {\bibinfo  {journal} {Nature
  Communications}\ }\textbf {\bibinfo {volume} {8}},\ \bibinfo {pages} {14176}
  (\bibinfo {year} {2017})}\BibitemShut {NoStop}%
\bibitem [{\citenamefont {Isobe}\ \emph {et~al.}(2020)\citenamefont {Isobe},
  \citenamefont {Xu},\ and\ \citenamefont {Fu}}]{Isobeeaay2497}%
  \BibitemOpen
  \bibfield  {author} {\bibinfo {author} {\bibfnamefont {H.}~\bibnamefont
  {Isobe}}, \bibinfo {author} {\bibfnamefont {S.-Y.}\ \bibnamefont {Xu}}, \
  and\ \bibinfo {author} {\bibfnamefont {L.}~\bibnamefont {Fu}},\ }\href
  {\doibase 10.1126/sciadv.aay2497} {\bibfield  {journal} {\bibinfo  {journal}
  {Science Advances}\ }\textbf {\bibinfo {volume} {6}} (\bibinfo {year}
  {2020}),\ 10.1126/sciadv.aay2497}\BibitemShut {NoStop}%
\bibitem [{\citenamefont {Zhang}\ \emph {et~al.}(1992)\citenamefont {Zhang},
  \citenamefont {Ma}, \citenamefont {Jin}, \citenamefont {Lu}, \citenamefont
  {Boden}, \citenamefont {Phelps}, \citenamefont {Stewart},\ and\ \citenamefont
  {Yakymyshyn}}]{doi:10.1063/1.107968}%
  \BibitemOpen
  \bibfield  {author} {\bibinfo {author} {\bibfnamefont {X.}~\bibnamefont
  {Zhang}}, \bibinfo {author} {\bibfnamefont {X.~F.}\ \bibnamefont {Ma}},
  \bibinfo {author} {\bibfnamefont {Y.}~\bibnamefont {Jin}}, \bibinfo {author}
  {\bibfnamefont {T.}~\bibnamefont {Lu}}, \bibinfo {author} {\bibfnamefont
  {E.~P.}\ \bibnamefont {Boden}}, \bibinfo {author} {\bibfnamefont {P.~D.}\
  \bibnamefont {Phelps}}, \bibinfo {author} {\bibfnamefont {K.~R.}\
  \bibnamefont {Stewart}}, \ and\ \bibinfo {author} {\bibfnamefont {C.~P.}\
  \bibnamefont {Yakymyshyn}},\ }\href {\doibase 10.1063/1.107968} {\bibfield
  {journal} {\bibinfo  {journal} {Applied Physics Letters}\ }\textbf {\bibinfo
  {volume} {61}},\ \bibinfo {pages} {3080} (\bibinfo {year} {1992})},\ \Eprint
  {http://arxiv.org/abs/https://doi.org/10.1063/1.107968}
  {https://doi.org/10.1063/1.107968} \BibitemShut {NoStop}%
\bibitem [{\citenamefont {Morimoto}\ and\ \citenamefont
  {Nagaosa}(2018)}]{Morimoto2018}%
  \BibitemOpen
  \bibfield  {author} {\bibinfo {author} {\bibfnamefont {T.}~\bibnamefont
  {Morimoto}}\ and\ \bibinfo {author} {\bibfnamefont {N.}~\bibnamefont
  {Nagaosa}},\ }\href {\doibase 10.1038/s41598-018-20539-2} {\bibfield
  {journal} {\bibinfo  {journal} {Scientific Reports}\ }\textbf {\bibinfo
  {volume} {8}},\ \bibinfo {pages} {2973} (\bibinfo {year} {2018})}\BibitemShut
  {NoStop}%
\bibitem [{\citenamefont {Tokura}\ and\ \citenamefont
  {Nagaosa}(2018)}]{Tokura2018}%
  \BibitemOpen
  \bibfield  {author} {\bibinfo {author} {\bibfnamefont {Y.}~\bibnamefont
  {Tokura}}\ and\ \bibinfo {author} {\bibfnamefont {N.}~\bibnamefont
  {Nagaosa}},\ }\href {\doibase 10.1038/s41467-018-05759-4} {\bibfield
  {journal} {\bibinfo  {journal} {Nature Communications}\ }\textbf {\bibinfo
  {volume} {9}},\ \bibinfo {pages} {3740} (\bibinfo {year} {2018})}\BibitemShut
  {NoStop}%
\bibitem [{\citenamefont {Nakai}\ and\ \citenamefont
  {Nagaosa}(2019)}]{PhysRevB.99.115201}%
  \BibitemOpen
  \bibfield  {author} {\bibinfo {author} {\bibfnamefont {R.}~\bibnamefont
  {Nakai}}\ and\ \bibinfo {author} {\bibfnamefont {N.}~\bibnamefont
  {Nagaosa}},\ }\href {\doibase 10.1103/PhysRevB.99.115201} {\bibfield
  {journal} {\bibinfo  {journal} {Phys. Rev. B}\ }\textbf {\bibinfo {volume}
  {99}},\ \bibinfo {pages} {115201} (\bibinfo {year} {2019})}\BibitemShut
  {NoStop}%
\bibitem [{\citenamefont {Wakatsuki}\ \emph {et~al.}(2017)\citenamefont
  {Wakatsuki}, \citenamefont {Saito}, \citenamefont {Hoshino}, \citenamefont
  {Itahashi}, \citenamefont {Ideue}, \citenamefont {Ezawa}, \citenamefont
  {Iwasa},\ and\ \citenamefont {Nagaosa}}]{Wakatsukie1602390}%
  \BibitemOpen
  \bibfield  {author} {\bibinfo {author} {\bibfnamefont {R.}~\bibnamefont
  {Wakatsuki}}, \bibinfo {author} {\bibfnamefont {Y.}~\bibnamefont {Saito}},
  \bibinfo {author} {\bibfnamefont {S.}~\bibnamefont {Hoshino}}, \bibinfo
  {author} {\bibfnamefont {Y.~M.}\ \bibnamefont {Itahashi}}, \bibinfo {author}
  {\bibfnamefont {T.}~\bibnamefont {Ideue}}, \bibinfo {author} {\bibfnamefont
  {M.}~\bibnamefont {Ezawa}}, \bibinfo {author} {\bibfnamefont
  {Y.}~\bibnamefont {Iwasa}}, \ and\ \bibinfo {author} {\bibfnamefont
  {N.}~\bibnamefont {Nagaosa}},\ }\href {\doibase 10.1126/sciadv.1602390}
  {\bibfield  {journal} {\bibinfo  {journal} {Science Advances}\ }\textbf
  {\bibinfo {volume} {3}} (\bibinfo {year} {2017}),\
  10.1126/sciadv.1602390}\BibitemShut {NoStop}%
\bibitem [{\citenamefont {Itahashi}\ \emph {et~al.}(2020)\citenamefont
  {Itahashi}, \citenamefont {Ideue}, \citenamefont {Saito}, \citenamefont
  {Shimizu}, \citenamefont {Ouchi}, \citenamefont {Nojima},\ and\ \citenamefont
  {Iwasa}}]{Itahashieaay9120}%
  \BibitemOpen
  \bibfield  {author} {\bibinfo {author} {\bibfnamefont {Y.~M.}\ \bibnamefont
  {Itahashi}}, \bibinfo {author} {\bibfnamefont {T.}~\bibnamefont {Ideue}},
  \bibinfo {author} {\bibfnamefont {Y.}~\bibnamefont {Saito}}, \bibinfo
  {author} {\bibfnamefont {S.}~\bibnamefont {Shimizu}}, \bibinfo {author}
  {\bibfnamefont {T.}~\bibnamefont {Ouchi}}, \bibinfo {author} {\bibfnamefont
  {T.}~\bibnamefont {Nojima}}, \ and\ \bibinfo {author} {\bibfnamefont
  {Y.}~\bibnamefont {Iwasa}},\ }\href {\doibase 10.1126/sciadv.aay9120}
  {\bibfield  {journal} {\bibinfo  {journal} {Science Advances}\ }\textbf
  {\bibinfo {volume} {6}} (\bibinfo {year} {2020}),\
  10.1126/sciadv.aay9120}\BibitemShut {NoStop}%
\bibitem [{\citenamefont {Ando}\ \emph {et~al.}(2020)\citenamefont {Ando},
  \citenamefont {Miyasaka}, \citenamefont {Li}, \citenamefont {Ishizuka},
  \citenamefont {Arakawa}, \citenamefont {Shiota}, \citenamefont {Moriyama},
  \citenamefont {Yanase},\ and\ \citenamefont {Ono}}]{Ando2020}%
  \BibitemOpen
  \bibfield  {author} {\bibinfo {author} {\bibfnamefont {F.}~\bibnamefont
  {Ando}}, \bibinfo {author} {\bibfnamefont {Y.}~\bibnamefont {Miyasaka}},
  \bibinfo {author} {\bibfnamefont {T.}~\bibnamefont {Li}}, \bibinfo {author}
  {\bibfnamefont {J.}~\bibnamefont {Ishizuka}}, \bibinfo {author}
  {\bibfnamefont {T.}~\bibnamefont {Arakawa}}, \bibinfo {author} {\bibfnamefont
  {Y.}~\bibnamefont {Shiota}}, \bibinfo {author} {\bibfnamefont
  {T.}~\bibnamefont {Moriyama}}, \bibinfo {author} {\bibfnamefont
  {Y.}~\bibnamefont {Yanase}}, \ and\ \bibinfo {author} {\bibfnamefont
  {T.}~\bibnamefont {Ono}},\ }\href {\doibase 10.1038/s41586-020-2590-4}
  {\bibfield  {journal} {\bibinfo  {journal} {Nature}\ }\textbf {\bibinfo
  {volume} {584}},\ \bibinfo {pages} {373} (\bibinfo {year}
  {2020})}\BibitemShut {NoStop}%
\bibitem [{\citenamefont {Yuan}\ and\ \citenamefont
  {Fu}(2021)}]{yuan2021supercurrent}%
  \BibitemOpen
  \bibfield  {author} {\bibinfo {author} {\bibfnamefont {N.~F.~Q.}\
  \bibnamefont {Yuan}}\ and\ \bibinfo {author} {\bibfnamefont {L.}~\bibnamefont
  {Fu}},\ }\href@noop {} {\enquote {\bibinfo {title} {Supercurrent diode effect
  and finite momentum superconductivity},}\ } (\bibinfo {year} {2021}),\
  \Eprint {http://arxiv.org/abs/2106.01909} {arXiv:2106.01909
  [cond-mat.supr-con]} \BibitemShut {NoStop}%
\bibitem [{\citenamefont {Daido}\ \emph {et~al.}(2021)\citenamefont {Daido},
  \citenamefont {Ikeda},\ and\ \citenamefont {Yanase}}]{daido2021intrinsic}%
  \BibitemOpen
  \bibfield  {author} {\bibinfo {author} {\bibfnamefont {A.}~\bibnamefont
  {Daido}}, \bibinfo {author} {\bibfnamefont {Y.}~\bibnamefont {Ikeda}}, \ and\
  \bibinfo {author} {\bibfnamefont {Y.}~\bibnamefont {Yanase}},\ }\href@noop {}
  {\enquote {\bibinfo {title} {Intrinsic superconducting diode effect},}\ }
  (\bibinfo {year} {2021}),\ \Eprint {http://arxiv.org/abs/2106.03326}
  {arXiv:2106.03326 [cond-mat.supr-con]} \BibitemShut {NoStop}%
\bibitem [{\citenamefont {He}\ \emph {et~al.}(2021)\citenamefont {He},
  \citenamefont {Tanaka},\ and\ \citenamefont
  {Nagaosa}}]{he2021phenomenological}%
  \BibitemOpen
  \bibfield  {author} {\bibinfo {author} {\bibfnamefont {J.~J.}\ \bibnamefont
  {He}}, \bibinfo {author} {\bibfnamefont {Y.}~\bibnamefont {Tanaka}}, \ and\
  \bibinfo {author} {\bibfnamefont {N.}~\bibnamefont {Nagaosa}},\ }\href@noop
  {} {\enquote {\bibinfo {title} {A phenomenological theory of superconductor
  diodes},}\ } (\bibinfo {year} {2021}),\ \Eprint
  {http://arxiv.org/abs/2106.03575} {arXiv:2106.03575 [cond-mat.supr-con]}
  \BibitemShut {NoStop}%
\bibitem [{\citenamefont {Du}\ \emph {et~al.}(2019)\citenamefont {Du},
  \citenamefont {Wang}, \citenamefont {Li}, \citenamefont {Lu},\ and\
  \citenamefont {Xie}}]{Du2019}%
  \BibitemOpen
  \bibfield  {author} {\bibinfo {author} {\bibfnamefont {Z.~Z.}\ \bibnamefont
  {Du}}, \bibinfo {author} {\bibfnamefont {C.~M.}\ \bibnamefont {Wang}},
  \bibinfo {author} {\bibfnamefont {S.}~\bibnamefont {Li}}, \bibinfo {author}
  {\bibfnamefont {H.-Z.}\ \bibnamefont {Lu}}, \ and\ \bibinfo {author}
  {\bibfnamefont {X.~C.}\ \bibnamefont {Xie}},\ }\href {\doibase
  10.1038/s41467-019-10941-3} {\bibfield  {journal} {\bibinfo  {journal}
  {Nature Communications}\ }\textbf {\bibinfo {volume} {10}},\ \bibinfo {pages}
  {3047} (\bibinfo {year} {2019})}\BibitemShut {NoStop}%
\bibitem [{\citenamefont {Nandy}\ and\ \citenamefont
  {Sodemann}(2019)}]{PhysRevB.100.195117}%
  \BibitemOpen
  \bibfield  {author} {\bibinfo {author} {\bibfnamefont {S.}~\bibnamefont
  {Nandy}}\ and\ \bibinfo {author} {\bibfnamefont {I.}~\bibnamefont
  {Sodemann}},\ }\href {\doibase 10.1103/PhysRevB.100.195117} {\bibfield
  {journal} {\bibinfo  {journal} {Phys. Rev. B}\ }\textbf {\bibinfo {volume}
  {100}},\ \bibinfo {pages} {195117} (\bibinfo {year} {2019})}\BibitemShut
  {NoStop}%
\bibitem [{\citenamefont {Du}\ \emph {et~al.}(2021)\citenamefont {Du},
  \citenamefont {Wang}, \citenamefont {Sun}, \citenamefont {Lu},\ and\
  \citenamefont {Xie}}]{Du2021}%
  \BibitemOpen
  \bibfield  {author} {\bibinfo {author} {\bibfnamefont {Z.~Z.}\ \bibnamefont
  {Du}}, \bibinfo {author} {\bibfnamefont {C.~M.}\ \bibnamefont {Wang}},
  \bibinfo {author} {\bibfnamefont {H.-P.}\ \bibnamefont {Sun}}, \bibinfo
  {author} {\bibfnamefont {H.-Z.}\ \bibnamefont {Lu}}, \ and\ \bibinfo {author}
  {\bibfnamefont {X.~C.}\ \bibnamefont {Xie}},\ }\href {\doibase
  10.1038/s41467-021-25273-4} {\bibfield  {journal} {\bibinfo  {journal}
  {Nature Communications}\ }\textbf {\bibinfo {volume} {12}},\ \bibinfo {pages}
  {5038} (\bibinfo {year} {2021})}\BibitemShut {NoStop}%
\bibitem [{\citenamefont {Morimoto}\ \emph {et~al.}(2016)\citenamefont
  {Morimoto}, \citenamefont {Zhong}, \citenamefont {Orenstein},\ and\
  \citenamefont {Moore}}]{PhysRevB.94.245121}%
  \BibitemOpen
  \bibfield  {author} {\bibinfo {author} {\bibfnamefont {T.}~\bibnamefont
  {Morimoto}}, \bibinfo {author} {\bibfnamefont {S.}~\bibnamefont {Zhong}},
  \bibinfo {author} {\bibfnamefont {J.}~\bibnamefont {Orenstein}}, \ and\
  \bibinfo {author} {\bibfnamefont {J.~E.}\ \bibnamefont {Moore}},\ }\href
  {\doibase 10.1103/PhysRevB.94.245121} {\bibfield  {journal} {\bibinfo
  {journal} {Phys. Rev. B}\ }\textbf {\bibinfo {volume} {94}},\ \bibinfo
  {pages} {245121} (\bibinfo {year} {2016})}\BibitemShut {NoStop}%
\bibitem [{\citenamefont {Ventura}\ \emph {et~al.}(2017)\citenamefont
  {Ventura}, \citenamefont {Passos}, \citenamefont {Lopes~dos Santos},
  \citenamefont {Viana Parente~Lopes},\ and\ \citenamefont
  {Peres}}]{PhysRevB.96.035431}%
  \BibitemOpen
  \bibfield  {author} {\bibinfo {author} {\bibfnamefont {G.~B.}\ \bibnamefont
  {Ventura}}, \bibinfo {author} {\bibfnamefont {D.~J.}\ \bibnamefont {Passos}},
  \bibinfo {author} {\bibfnamefont {J.~M.~B.}\ \bibnamefont {Lopes~dos
  Santos}}, \bibinfo {author} {\bibfnamefont {J.~M.}\ \bibnamefont {Viana
  Parente~Lopes}}, \ and\ \bibinfo {author} {\bibfnamefont {N.~M.~R.}\
  \bibnamefont {Peres}},\ }\href {\doibase 10.1103/PhysRevB.96.035431}
  {\bibfield  {journal} {\bibinfo  {journal} {Phys. Rev. B}\ }\textbf {\bibinfo
  {volume} {96}},\ \bibinfo {pages} {035431} (\bibinfo {year}
  {2017})}\BibitemShut {NoStop}%
\bibitem [{\citenamefont {Passos}\ \emph {et~al.}(2018)\citenamefont {Passos},
  \citenamefont {Ventura}, \citenamefont {Lopes}, \citenamefont {Santos},\ and\
  \citenamefont {Peres}}]{PhysRevB.97.235446}%
  \BibitemOpen
  \bibfield  {author} {\bibinfo {author} {\bibfnamefont {D.~J.}\ \bibnamefont
  {Passos}}, \bibinfo {author} {\bibfnamefont {G.~B.}\ \bibnamefont {Ventura}},
  \bibinfo {author} {\bibfnamefont {J.~M. V.~P.}\ \bibnamefont {Lopes}},
  \bibinfo {author} {\bibfnamefont {J.~M. B. L.~d.}\ \bibnamefont {Santos}}, \
  and\ \bibinfo {author} {\bibfnamefont {N.~M.~R.}\ \bibnamefont {Peres}},\
  }\href {\doibase 10.1103/PhysRevB.97.235446} {\bibfield  {journal} {\bibinfo
  {journal} {Phys. Rev. B}\ }\textbf {\bibinfo {volume} {97}},\ \bibinfo
  {pages} {235446} (\bibinfo {year} {2018})}\BibitemShut {NoStop}%
\bibitem [{\citenamefont {Watanabe}\ and\ \citenamefont
  {Yanase}(2021)}]{PhysRevX.11.011001}%
  \BibitemOpen
  \bibfield  {author} {\bibinfo {author} {\bibfnamefont {H.}~\bibnamefont
  {Watanabe}}\ and\ \bibinfo {author} {\bibfnamefont {Y.}~\bibnamefont
  {Yanase}},\ }\href {\doibase 10.1103/PhysRevX.11.011001} {\bibfield
  {journal} {\bibinfo  {journal} {Phys. Rev. X}\ }\textbf {\bibinfo {volume}
  {11}},\ \bibinfo {pages} {011001} (\bibinfo {year} {2021})}\BibitemShut
  {NoStop}%
\bibitem [{\citenamefont {Watanabe}\ \emph
  {et~al.}(2021{\natexlab{a}})\citenamefont {Watanabe}, \citenamefont {Daido},\
  and\ \citenamefont {Yanase}}]{watanabe2021nonreciprocal1}%
  \BibitemOpen
  \bibfield  {author} {\bibinfo {author} {\bibfnamefont {H.}~\bibnamefont
  {Watanabe}}, \bibinfo {author} {\bibfnamefont {A.}~\bibnamefont {Daido}}, \
  and\ \bibinfo {author} {\bibfnamefont {Y.}~\bibnamefont {Yanase}},\
  }\href@noop {} {\enquote {\bibinfo {title} {Nonreciprocal optical response in
  parity-breaking superconductors},}\ } (\bibinfo {year}
  {2021}{\natexlab{a}}),\ \Eprint {http://arxiv.org/abs/2109.14866}
  {arXiv:2109.14866 [cond-mat.supr-con]} \BibitemShut {NoStop}%
\bibitem [{\citenamefont {Watanabe}\ \emph
  {et~al.}(2021{\natexlab{b}})\citenamefont {Watanabe}, \citenamefont {Daido},\
  and\ \citenamefont {Yanase}}]{watanabe2021nonreciprocal2}%
  \BibitemOpen
  \bibfield  {author} {\bibinfo {author} {\bibfnamefont {H.}~\bibnamefont
  {Watanabe}}, \bibinfo {author} {\bibfnamefont {A.}~\bibnamefont {Daido}}, \
  and\ \bibinfo {author} {\bibfnamefont {Y.}~\bibnamefont {Yanase}},\
  }\href@noop {} {\enquote {\bibinfo {title} {Nonreciprocal meissner response
  in parity-mixed superconductors},}\ } (\bibinfo {year}
  {2021}{\natexlab{b}}),\ \Eprint {http://arxiv.org/abs/2109.14874}
  {arXiv:2109.14874 [cond-mat.supr-con]} \BibitemShut {NoStop}%
\bibitem [{\citenamefont {Michishita}\ and\ \citenamefont
  {Peters}(2021)}]{PhysRevB.103.195133}%
  \BibitemOpen
  \bibfield  {author} {\bibinfo {author} {\bibfnamefont {Y.}~\bibnamefont
  {Michishita}}\ and\ \bibinfo {author} {\bibfnamefont {R.}~\bibnamefont
  {Peters}},\ }\href {\doibase 10.1103/PhysRevB.103.195133} {\bibfield
  {journal} {\bibinfo  {journal} {Phys. Rev. B}\ }\textbf {\bibinfo {volume}
  {103}},\ \bibinfo {pages} {195133} (\bibinfo {year} {2021})}\BibitemShut
  {NoStop}%
\bibitem [{\citenamefont {Gao}\ \emph {et~al.}(2014)\citenamefont {Gao},
  \citenamefont {Yang},\ and\ \citenamefont {Niu}}]{PhysRevLett.112.166601}%
  \BibitemOpen
  \bibfield  {author} {\bibinfo {author} {\bibfnamefont {Y.}~\bibnamefont
  {Gao}}, \bibinfo {author} {\bibfnamefont {S.~A.}\ \bibnamefont {Yang}}, \
  and\ \bibinfo {author} {\bibfnamefont {Q.}~\bibnamefont {Niu}},\ }\href
  {\doibase 10.1103/PhysRevLett.112.166601} {\bibfield  {journal} {\bibinfo
  {journal} {Phys. Rev. Lett.}\ }\textbf {\bibinfo {volume} {112}},\ \bibinfo
  {pages} {166601} (\bibinfo {year} {2014})}\BibitemShut {NoStop}%
\bibitem [{\citenamefont {Gao}\ and\ \citenamefont
  {Xiao}(2019)}]{PhysRevLett.122.227402}%
  \BibitemOpen
  \bibfield  {author} {\bibinfo {author} {\bibfnamefont {Y.}~\bibnamefont
  {Gao}}\ and\ \bibinfo {author} {\bibfnamefont {D.}~\bibnamefont {Xiao}},\
  }\href {\doibase 10.1103/PhysRevLett.122.227402} {\bibfield  {journal}
  {\bibinfo  {journal} {Phys. Rev. Lett.}\ }\textbf {\bibinfo {volume} {122}},\
  \bibinfo {pages} {227402} (\bibinfo {year} {2019})}\BibitemShut {NoStop}%
\bibitem [{\citenamefont {Parker}\ \emph {et~al.}(2019)\citenamefont {Parker},
  \citenamefont {Morimoto}, \citenamefont {Orenstein},\ and\ \citenamefont
  {Moore}}]{PhysRevB.99.045121}%
  \BibitemOpen
  \bibfield  {author} {\bibinfo {author} {\bibfnamefont {D.~E.}\ \bibnamefont
  {Parker}}, \bibinfo {author} {\bibfnamefont {T.}~\bibnamefont {Morimoto}},
  \bibinfo {author} {\bibfnamefont {J.}~\bibnamefont {Orenstein}}, \ and\
  \bibinfo {author} {\bibfnamefont {J.~E.}\ \bibnamefont {Moore}},\ }\href
  {\doibase 10.1103/PhysRevB.99.045121} {\bibfield  {journal} {\bibinfo
  {journal} {Phys. Rev. B}\ }\textbf {\bibinfo {volume} {99}},\ \bibinfo
  {pages} {045121} (\bibinfo {year} {2019})}\BibitemShut {NoStop}%
\bibitem [{Note1()}]{Note1}%
  \BibitemOpen
  \bibinfo {note} {We can see $\sigma _M\propto \eta $ from
  $(G^R_n-G^A_n)_{i\omega _M}=2i\eta /((i\omega _M-\epsilon _n)^2+4\eta ^2)$.
  We can also see $\sigma _M\propto \eta $ in figure~\ref
  {fig:mudep}.}\BibitemShut {Stop}%
\bibitem [{\citenamefont {Sipe}\ and\ \citenamefont
  {Shkrebtii}(2000)}]{PhysRevB.61.5337}%
  \BibitemOpen
  \bibfield  {author} {\bibinfo {author} {\bibfnamefont {J.~E.}\ \bibnamefont
  {Sipe}}\ and\ \bibinfo {author} {\bibfnamefont {A.~I.}\ \bibnamefont
  {Shkrebtii}},\ }\href {\doibase 10.1103/PhysRevB.61.5337} {\bibfield
  {journal} {\bibinfo  {journal} {Phys. Rev. B}\ }\textbf {\bibinfo {volume}
  {61}},\ \bibinfo {pages} {5337} (\bibinfo {year} {2000})}\BibitemShut
  {NoStop}%
\bibitem [{\citenamefont {Ahn}\ \emph {et~al.}(2020)\citenamefont {Ahn},
  \citenamefont {Guo},\ and\ \citenamefont {Nagaosa}}]{PhysRevX.10.041041}%
  \BibitemOpen
  \bibfield  {author} {\bibinfo {author} {\bibfnamefont {J.}~\bibnamefont
  {Ahn}}, \bibinfo {author} {\bibfnamefont {G.-Y.}\ \bibnamefont {Guo}}, \ and\
  \bibinfo {author} {\bibfnamefont {N.}~\bibnamefont {Nagaosa}},\ }\href
  {\doibase 10.1103/PhysRevX.10.041041} {\bibfield  {journal} {\bibinfo
  {journal} {Phys. Rev. X}\ }\textbf {\bibinfo {volume} {10}},\ \bibinfo
  {pages} {041041} (\bibinfo {year} {2020})}\BibitemShut {NoStop}%
\bibitem [{Note2()}]{Note2}%
  \BibitemOpen
  \bibinfo {note} {One can check $\Omega ^{\alpha ,\beta \gamma }_{n,m} =
  (\Omega ^{\alpha \beta }_{n,m}\protect \mathcal {J}^{\gamma }_m+(\beta
  \leftrightarrow \gamma ))/\epsilon _{nm}$ by substituting $l=m$ in Eq.~(\ref
  {gBC_f})}\BibitemShut {NoStop}%
\bibitem [{\citenamefont {Wilson}\ and\ \citenamefont
  {Yoffe}(1969)}]{doi:10.1080/00018736900101307}%
  \BibitemOpen
  \bibfield  {author} {\bibinfo {author} {\bibfnamefont {J.}~\bibnamefont
  {Wilson}}\ and\ \bibinfo {author} {\bibfnamefont {A.}~\bibnamefont {Yoffe}},\
  }\href {\doibase 10.1080/00018736900101307} {\bibfield  {journal} {\bibinfo
  {journal} {Advances in Physics}\ }\textbf {\bibinfo {volume} {18}},\ \bibinfo
  {pages} {193} (\bibinfo {year} {1969})},\ \Eprint
  {http://arxiv.org/abs/https://doi.org/10.1080/00018736900101307}
  {https://doi.org/10.1080/00018736900101307} \BibitemShut {NoStop}%
\bibitem [{\citenamefont {Liu}\ \emph {et~al.}(2015)\citenamefont {Liu},
  \citenamefont {Xiao}, \citenamefont {Yao}, \citenamefont {Xu},\ and\
  \citenamefont {Yao}}]{C4CS00301B}%
  \BibitemOpen
  \bibfield  {author} {\bibinfo {author} {\bibfnamefont {G.-B.}\ \bibnamefont
  {Liu}}, \bibinfo {author} {\bibfnamefont {D.}~\bibnamefont {Xiao}}, \bibinfo
  {author} {\bibfnamefont {Y.}~\bibnamefont {Yao}}, \bibinfo {author}
  {\bibfnamefont {X.}~\bibnamefont {Xu}}, \ and\ \bibinfo {author}
  {\bibfnamefont {W.}~\bibnamefont {Yao}},\ }\href {\doibase
  10.1039/C4CS00301B} {\bibfield  {journal} {\bibinfo  {journal} {Chem. Soc.
  Rev.}\ }\textbf {\bibinfo {volume} {44}},\ \bibinfo {pages} {2643} (\bibinfo
  {year} {2015})}\BibitemShut {NoStop}%
\bibitem [{\citenamefont {Kanasugi}\ and\ \citenamefont
  {Yanase}(2020)}]{PhysRevB.102.094507}%
  \BibitemOpen
  \bibfield  {author} {\bibinfo {author} {\bibfnamefont {S.}~\bibnamefont
  {Kanasugi}}\ and\ \bibinfo {author} {\bibfnamefont {Y.}~\bibnamefont
  {Yanase}},\ }\href {\doibase 10.1103/PhysRevB.102.094507} {\bibfield
  {journal} {\bibinfo  {journal} {Phys. Rev. B}\ }\textbf {\bibinfo {volume}
  {102}},\ \bibinfo {pages} {094507} (\bibinfo {year} {2020})}\BibitemShut
  {NoStop}%
\bibitem [{\citenamefont {Bezanson}\ \emph {et~al.}(2017)\citenamefont
  {Bezanson}, \citenamefont {Edelman}, \citenamefont {Karpinski},\ and\
  \citenamefont {Shah}}]{bezanson2017julia}%
  \BibitemOpen
  \bibfield  {author} {\bibinfo {author} {\bibfnamefont {J.}~\bibnamefont
  {Bezanson}}, \bibinfo {author} {\bibfnamefont {A.}~\bibnamefont {Edelman}},
  \bibinfo {author} {\bibfnamefont {S.}~\bibnamefont {Karpinski}}, \ and\
  \bibinfo {author} {\bibfnamefont {V.~B.}\ \bibnamefont {Shah}},\ }\href
  {https://doi.org/10.1137/141000671} {\bibfield  {journal} {\bibinfo
  {journal} {SIAM review}\ }\textbf {\bibinfo {volume} {59}},\ \bibinfo {pages}
  {65} (\bibinfo {year} {2017})}\BibitemShut {NoStop}%
\bibitem [{\citenamefont {Matsushita}\ \emph {et~al.}(2020)\citenamefont
  {Matsushita}, \citenamefont {Fujimoto},\ and\ \citenamefont
  {Schnyder}}]{PhysRevResearch.2.043311}%
  \BibitemOpen
  \bibfield  {author} {\bibinfo {author} {\bibfnamefont {T.}~\bibnamefont
  {Matsushita}}, \bibinfo {author} {\bibfnamefont {S.}~\bibnamefont
  {Fujimoto}}, \ and\ \bibinfo {author} {\bibfnamefont {A.~P.}\ \bibnamefont
  {Schnyder}},\ }\href {\doibase 10.1103/PhysRevResearch.2.043311} {\bibfield
  {journal} {\bibinfo  {journal} {Phys. Rev. Research}\ }\textbf {\bibinfo
  {volume} {2}},\ \bibinfo {pages} {043311} (\bibinfo {year}
  {2020})}\BibitemShut {NoStop}%
\bibitem [{\citenamefont {Yamakage}\ \emph {et~al.}(2016)\citenamefont
  {Yamakage}, \citenamefont {Yamakawa}, \citenamefont {Tanaka},\ and\
  \citenamefont {Okamoto}}]{doi:10.7566/JPSJ.85.013708}%
  \BibitemOpen
  \bibfield  {author} {\bibinfo {author} {\bibfnamefont {A.}~\bibnamefont
  {Yamakage}}, \bibinfo {author} {\bibfnamefont {Y.}~\bibnamefont {Yamakawa}},
  \bibinfo {author} {\bibfnamefont {Y.}~\bibnamefont {Tanaka}}, \ and\ \bibinfo
  {author} {\bibfnamefont {Y.}~\bibnamefont {Okamoto}},\ }\href {\doibase
  10.7566/JPSJ.85.013708} {\bibfield  {journal} {\bibinfo  {journal} {Journal
  of the Physical Society of Japan}\ }\textbf {\bibinfo {volume} {85}},\
  \bibinfo {pages} {013708} (\bibinfo {year} {2016})},\ \Eprint
  {http://arxiv.org/abs/https://doi.org/10.7566/JPSJ.85.013708}
  {https://doi.org/10.7566/JPSJ.85.013708} \BibitemShut {NoStop}%
\bibitem [{\citenamefont {Chan}\ \emph {et~al.}(2016)\citenamefont {Chan},
  \citenamefont {Chiu}, \citenamefont {Chou},\ and\ \citenamefont
  {Schnyder}}]{PhysRevB.93.205132}%
  \BibitemOpen
  \bibfield  {author} {\bibinfo {author} {\bibfnamefont {Y.-H.}\ \bibnamefont
  {Chan}}, \bibinfo {author} {\bibfnamefont {C.-K.}\ \bibnamefont {Chiu}},
  \bibinfo {author} {\bibfnamefont {M.~Y.}\ \bibnamefont {Chou}}, \ and\
  \bibinfo {author} {\bibfnamefont {A.~P.}\ \bibnamefont {Schnyder}},\ }\href
  {\doibase 10.1103/PhysRevB.93.205132} {\bibfield  {journal} {\bibinfo
  {journal} {Phys. Rev. B}\ }\textbf {\bibinfo {volume} {93}},\ \bibinfo
  {pages} {205132} (\bibinfo {year} {2016})}\BibitemShut {NoStop}%
\end{thebibliography}%

\end{document}